\documentclass[12pt]{iopart}

\usepackage{amsmath}   
\usepackage{booktabs}
\usepackage{colortbl}  
\usepackage{xcolor}    
\usepackage{caption}
\usepackage{graphicx}
\usepackage[
  backend=biber,
  sorting=none
]{biblatex}
\usepackage[normalem]{ulem}  

\definecolor{lightgray}{gray}{0.9}
\definecolor{lightblue}{rgb}{0.8, 0.85, 1}
\definecolor{saddlebrown}{RGB}{139,69,19}
\definecolor{softred}{RGB}{220, 80, 80}
\definecolor{sage}{RGB}{220,230,220}
\definecolor{emerald}{RGB}{0,150,0}
\definecolor{sagegreen}{RGB}{120,170,140}

\begin{document}

\title[]{Core-edge integrated modeling of ARC: on the effect of impurity transport and detachment conditions.}

\author{M. Muraca$^1$, P. Rodriguez-Fernandez$^1$, N.T. Howard$^1$, J. Hall$^1$, G. Tardini$^2$, D. Silvagni$^2$, T. Body$^3$, J. Hillesheim$^3$}

\address{$^1$MIT Plasma Science and Fusion Center, 167 Albany St, Cambridge, MA 02139, USA \\
$^2$Max Planck Institute for Plasma Physics, Boltzmannstrasse 2, Garching bei M\"unchen, 85748, Germany \\
$^3$Commonwealth Fusion Systems, 117 Hospital Rd, Devens, MA 01434, USA}
\ead{mmuraca@mit.edu}
\vspace{10pt}
\begin{indented}
\item[]February 2026
\end{indented}

\begin{abstract}
Integrated modeling of ARC H-modes has been conducted to assess the feasibility of high-performance scenarios compatible with divertor detachment. The analysis incorporates self-consistent evolution of impurity radiation and density profiles, demonstrating that fusion power levels approaching a GW can be achieved while maintaining divertor temperatures below 2 eV with Ar seeding. Sensitivity studies reveal a strong dependence of fusion power on the separatrix density, with performance spanning 750–1000 MW, and a weaker dependence on enrichment factor and pedestal density. Alternative seeding strategies using Neon have also been explored. Plasmas with Argon seeding consistently access H-mode, providing the highest fusion power and detached divertor operation, whereas Neon seeding leads to lower performance (600–850 MW) and less robust H-mode access, due to excessive core impurity accumulation. A small W impurity peaking has been found, with decreasing values at higher $Z_{eff}$. Further analyses incorporate reduced momentum transport modeling, and sensitivity studies of neoclassical impurity transport, confirming the robustness of the results. Overall, these findings support the viability of high-performing H-mode operation in ARC, ensuring divertor protection, enabled through Argon and Neon impurity seeding.
\end{abstract}

\section{Introduction}

The design of commercial power plants must satisfy stringent performance and respect specific engineering constraints, while operating in regimes which are only partially validated by experimental data. In this context, integrated transport modeling is essential to predict fusion performance, explore operational limits, and guide design optimization, providing the evolution of kinetic profiles, power fluxes and exhaust loads.

The ARC commercial power plant \cite{sorbom_arc_2015}, [Hillesheim JPP 2026, accepted], characterized by compact size and high magnetic field, shows a promising pathway toward high fusion power density. However, fusion performance predictions from integrated modeling are typically sensitive to both physics and engineering assumptions. In particular, core confinement, H-mode pedestals and exhaust handling depend nonlinearly on parameters like magnetic field and plasma current \cite{verdoolaege_updated_2021, reinke_heat_2017, angioni_dependence_2023}, density \cite{martin_power_2008, schmidtmayr_investigation_2018, delabie_empirical_2026}, shaping \cite{horton_dependence_2002, snyder_high_2019} and impurity content \cite{ fajardo_theorybased_2025, muraca_integrated_2025, ivanova-stanik_integrated_2023} \cite{muraca_impurity_2026}. An ideal approach to strengthen the reliability of predictions is to increase the fidelity of the simulations by including as much physics description as possible and reducing the number of assumptions. However, a completely self-consistent framework remains impractical, particularly with reasonable computational cost \cite{Falchetto2014Nucl.Fusion_European}, and a set of sensitivity studies is often required on input assumptions to quantify the robustness of the predictions.

In high-performance H-mode regimes, the pedestal supports higher kinetic profiles and plays a central role in determining the overall fusion power. Its structure arises from a complex interplay between fueling, transport, stability, and atomic physics, which remains only partially understood to date \cite{fenstermacher_progress_2025}. Example of outstanding gaps in pedestal formation are discussed in \cite{groebner_elements_2023, aiba_modeling_2025}. As a result, the prediction of pedestal height and width is prone to substantial uncertainties, making the self-consistent coupling of core and edge modeling challenging, but necessary to provide reliable results.

In fusion power plants, detachment \cite{kallenbach_impurity_2013, krasheninnikov_physics_2017, soukhanovskii_review_2017} will be needed to prevent excessive material erosion and melting of material surfaces. The seeding of impurities facilitates detachment, but can cause impurity core penetration, with consequent high radiation, fuel dilution and H-L transition \cite{maggi_experimental_2013}. Therefore, an accurate modeling of impurity transport, charge-state equilibrium, radiation and SOL physics is indispensable to identify integrated solutions which allow both divertor protection and high core performance.

Hence, the integrated modeling of fusion power plants requires a comprehensive approach in which core transport, pedestal physics, exhaust handling and impurity behavior are treated consistently. Such a framework, based on a set of reduced models, has been adopted in this work, predicting the fusion performance of H-mode plasmas and quantifying uncertainties with respect to the input parameters. Reduced models have been selected to optimize computational cost while preserving reliable results.

The remainder of the article is organized as follows: in section \ref{sec:simsetup} the setup of the simulations is described; in section \ref{sec:exploration} an exploration of ARC-class configurations has been introduced to explore the variation of fusion power with design parameters; in section \ref{sec:results} a set of sensitivity studies are analyzed, highlighting the physics parameters which mainly impact performance and showcasing different seeding strategies; in section 5 the conclusions are discussed.

\section{Simulation Setup}
\label{sec:simsetup}
To assess plasma performance, multiple physics models must be combined within an integrated framework capable of predicting kinetic profiles. In this work, such task has been carried out using the ASTRA transport solver \cite{pereverzev_astra_1991}. Although ASTRA is capable of simulating the time evolution of a discharge, we focus here exclusively on stationary flat-top phases. The equilibrium is computed with SPIDER \cite{ivanov_new_2005}, which performs fixed-boundary computations, taking as inputs current and pressure profiles self-consistently calculated in ASTRA, and the last closed flux surface (LCFS) coordinates, which are obtained with free-boundary FreeGS \cite{_freegsplasma_2024} simulations, external to ASTRA runs. Both SPIDER and FreeGS solve the Grad-Shafranov equation \cite{grad_hydromagnetic_1958, shafranov_plasma_1966}.

Sawteeth can modify safety factor and kinetic profiles in ARC plasmas, especially given the absence of non-inductive current drive. Hence, TRANSP simulations have been performed, using the Porcelli model \cite{porcelli_model_1996}, that includes the stabilization provided by fast ions, to compute inversion radius and sawtooth period. This model has been used with the same inputs as in \cite{rodriguez-fernandez_predictions_2020}. The resulting safety factor profile prior to the sawtooth crash, that exhibits on axis values around 0.75, has been prescribed in ASTRA simulations.

Turbulent transport in the core is calculated with the quasi-linear model TGLF \cite{staebler_verification_2021}, which computes turbulence associated with micro-instabilities such as Electron Temperature Gradient (ETG), Trapped Electron Mode (TEM), and Ion Temperature Gradient (ITG) modes \cite{weiland_collective_2000, dimits_comparisons_2000, garbet_introduction_2006}. TGLF has been extensively validated across several devices and plasma scenarios, demonstrating good predictive capability \cite{rodriguez-fernandez_predictfirst_2019, rodriguez-fernandez_perturbative_2019, angioni_confinement_2022, baiocchi_turbulent_2015, creely_validation_2017, staebler_quasilinear_2024}. In this paper, the main TGLF control settings have been chosen consistently with [Howard JPP 2026, accepted], including SAT2 saturation rule \cite{staebler_verification_2021}, electromagnetic effects (through $\delta A_\parallel$), five plasma species, Miller geometry \cite{stacey_representation_2009} and a maximum of six parallel basis functions, while the direct effect of fast ions on turbulence is not modeled, since in the present framework ICRH minority is treated as a thermal species and the evolution of $\alpha$ particles is not included. To provide separately diffusivity ($D$) and convection ($v$) for impurities, TGLF is called twice at each time step, with the same strategy described in \cite{fajardo_fullradius_2024} and \cite{muraca_impurity_2026}.

Neoclassical impurity transport is computed with FACIT \cite{fajardo_analytical_2022, fajardo_analytical_2023} up to the separatrix. This is a fast analytical model, which includes rotational effects and poloidal asymmetries, and calculates flux-surface-averaged transport coefficients for arbitrary mass, charge, collisionality and radial position in the confined region. Atomic processes like ionization and recombination are computed self-consistently in STRAHL \cite{dux_strahl_2006}, which is called by ASTRA at every time step. Moreover, for each impurity species, STRAHL calculates the line radiation and transport of each charge state, using as input diffusivity ($D$), convection ($v$) and a wall source. \

The pedestal pressure is predicted using a neural network (NN) trained on EPED simulations of ARC [Howard JPP 2026, accepted]. This model self-consistently calculates the pedestal height and width, as inputs like the stored energy and $Z_{eff}$ evolve. In EPED-NN, the pedestal density (i.e. $n_{ped}$) is an input parameter. $n_{ped}$ is imposed in the simulations, as it depends on an interplay between source and edge transport physics \cite{groebner_elements_2023, saarelma_density_2024}, whose modeling is beyond the scope of this work. Therefore, no D/T particle source model has been included and density scans have been performed across the paper to account for uncertainties. The impurity transport from the pedestal top to the separatrix is tuned to achieve specific pedestal concentrations. This simple approach is justified by the lack of robust edge transport models to be used in integrated modeling of H-modes. The effect of ELMs on the divertor, which can impact performance and exhaust \cite{eich_elm_2017}, is not modeled. Details on potential ELM-free scenarios for ARC can be found in [Eich JPP 2026, accepted].

ARC's auxiliary heating power will be provided by Ion Cyclotron Resonance Heating (ICRH) \cite{lin_physics_2020}. No model is yet present in ASTRA to simulate self-consistent ICRH absorption profiles. Therefore, external to ASTRA runs, simulations are performed using TRANSP \cite{pankin_transp_2024}, coupled with TORIC \cite{brambilla_numerical_1999} and FPPMOD \cite{hammett_fast_1986}, to describe RF wave propagation and Fokker-Planck collisions. This approach has also been adopted in \cite{muraca_integrated_2025} and \cite{muraca_impurity_2026}. The calculated deposition profiles are then imported into ASTRA simulations. Ohmic power, collisional energy exchange, fusion power and radiation are computed as in \cite{muraca_integrated_2025}, \cite{muraca_impurity_2026}.

The required SOL seeding concentration (i.e. $f_{seed,SOL}$) to access detachment has been computed with the extended Lengyel model \cite{body_simple_2025}, which also calculates the separatrix temperature (i.e. $T_{sep}=T_{i,sep}=T_{e,sep}$) and divertor neutral pressure (i.e. $p_{0,div}$), taking the density and heat flux at the separatrix (i.e. $n_{sep}$ and $P_{loss}$) as inputs, and assuming a few other machine-dependent geometrical parameters consistently with [Eich JPP 2026, accepted]. It is important to mention that while $P_{loss}$ is self-consistently computed in ASTRA through the integration of core and edge modeling, $n_{sep}$ has been assumed. In fact, predicting precisely this parameter is difficult with standard integrated modeling due to its dependence on divertor neutral pressure, recycling, and open field–line physics in the scrape-off layer \cite{manz_how_2025, silvagni_separatrix_2025}, requiring multi-dimensional treatment beyond the capabilities of many reduced models \cite{kaveeva_solpsiter_2023}. A divertor temperature of 2 eV has been assumed, roughly corresponding to a condition of 50\% momentum loss from the outer mid-plane to the divertor, to ensure detached conditions \cite{groth_characterisation_2023}, [Eich JPP 2026, accepted].

While the extended Lengyel (X-Lengyel) model can provide $f_{seed,SOL}$, enrichment factors must be assumed to estimate the concentration of impurities in the core. The enrichment used in this work is defined as $\epsilon=f_{seed,SOL}/f_{seed,core}$. Therefore, $\epsilon>1$ represents a low core impurity penetration. Unless otherwise specified, $\epsilon$ is obtained employing the scaling in figure 8 of \cite{kallenbach_divertor_2024}. The expression is
\begin{equation}
    \label{eq:Kallenbach_reg}
    \epsilon=41\cdot Z^{-0.5}\:p_{0,div}^{-0.4}\left(\frac{E_{ion,Z}}{E_{ion,D}}\right)^{-5.8},
\end{equation}
where $Z$ is the charge of the seeded impurity and $E_{ion,Z}$ and $E_{ion,D}$ are the first ionization energy of the seeded impurity and D.
In the present framework, the scaling factors have been computed to assign a top of pedestal concentration for the seeded species (i.e. $f_{seed,top}$), while the impurities are evolved self-consistently with the transport models from pedestal top to magnetic axis. $f_{seed,top}$ is obtained by tuning the turbulent transport coefficients in the pedestal region. The resulting $Z_{eff}$ at the pedestal (i.e. $Z_{eff,pedestal}$), which is obtained applying the tuned constant transport in the edge, has been used as input for pedestal predictions, according to the EPED implementation used to derive the NN.

The coupling of core-pedestal-SOL modeling is realized through feedback algorithms between X-Lengyel, EPED-NN and core models, by exchanging $P_{loss}$, $T_{sep}$, $f_{seed,SOL}$, $p_{0,div}$, $\epsilon$ and $Z_{eff,pedestal}$.

W and H are modeled to reflect respectively wall / divertor erosion and minority ions used for ICRH heating \cite{vaneester_minority_2012}. Since ASTRA lacks a self-consistent Fokker-Planck model, H is treated as thermal species, without addressing collisions with other species. The lack of fast H modeling should affect minimally the fusion power, because the heating and fast particles population is dominated by the $\alpha$ particles in the analyzed scenarios. Moreover, the ICRH minority is included only to provide realistic DT dilution and to ensure sufficient near-axis H concentrations for efficient ICRH absorption. While enrichment factors have been used for the seeded impurity, W and H concentrations at the pedestal top (i.e. $f_{W,top}$ and $f_{H,top}$) have been assumed equal to respectively $1.5\cdot10^{-5}$ and $0.05$. The edge transport coefficients are tuned to match $f_{W,top}=1.5\cdot10^{-5}$ for W, $f_{H,top}=0.05$ for H, and the value predicted from Eq. \ref{eq:Kallenbach_reg} for the seeded species. Simulations performed in [Howard JPP 2026, accepted] assumed $f_H=0.03$, which should be sufficient to absorb ICRH heating near-axis, but here a conservative approach has been chosen. W has a small effect on pedestal predictions, since it weakly affects $Z_{eff}$, due to its low concentration.

The transport has been computed for electrons, seeded impurity, H and W, while the main ion population, composed by a 50-50\% mix of D and T has been lumped into a species with charge 1 and mass 2.5, and its density has been assumed to fulfill quasi-neutrality. The He ash has not been included, given the uncertainties associated with its concentration and the limit of 5 species (i.e. electrons, lumped DT, Ar/Ne, W and H) imposed by the present TGLF implementation in ASTRA.

\section{Exploration of high-field FPP design points}
\label{sec:exploration}
Using the framework described in the previous section, several simulations have been executed, assuming a series of Fusion Pilot Plants (FPPs) with different fusion power targets, including self-consistent impurity transport modeling and detached divertor conditions. This collection of runs encompasses various transport studies conducted at MIT, performed to demonstrate key trends captured in the performance. The design points span different values of magnetic field, plasma current, shaping, minor radius, major radius and density, while limiting variations of $q_{95}$ between 4 and 4.9, $q_{Uckan}$ between 3.1 and 3.8, and aspect ratio between 3.4 and 3.9. $q_{95}$ is the safety factor at 95\% of the normalized poloidal magnetic flux and $q_{Uckan}$ is $\left( \frac{5a^2B_t}{R_0I_P}\right)\frac{1+k_{95}^2\left(1+2\delta_{95}^2-1.2\delta_{95}^3\right)}{2}$. 

The ranges explored for the design variables are listed in table \ref{tab:designs}, together with the nominal values of ARC V3A, which was introduced in [Hillesheim, JPP 2026, accepted].
\begin{table}[h]
  \centering
  \caption{Ranges of variation of parameters for different ARC-class designs. $\kappa_{sep}$ and $\delta_{sep}$ indicate elongation and triangularity at the separatrix. The third column shows the nominal parameters of the ARC V3A design [Hillesheim JPP 2026, accepted]}
  \begin{tabular}{>{\columncolor{lightgray}}c | c | c}
    \toprule
    \rowcolor{lightgray} 
    \textbf{Parameter} & \textbf{Range} & \textbf{ARC V3A} \\
    \midrule
          $B_t$ ($T$) &  [9.8 - 11.7]  &  \cellcolor{red!50!white}\textbf{11.4} \\
          $I_p$ ($MA$)  &  [10.0 - 13.8]  &  \cellcolor{red!50!white}\textbf{12.0} \\
          $\delta_{sep}$  &  [0.48 - 0.65]  &  \cellcolor{red!50!white}\textbf{0.65} \\
          $\kappa_{sep}$  &  [1.69 - 1.9]  &  \cellcolor{red!50!white}\textbf{1.8} \\
          $a$ ($m$)  &  [1.05 - 1.27]  &  \cellcolor{red!50!white}\textbf{1.18} \\
          $R$ ($m$)  &  [4.08 - 4.62]  &  \cellcolor{red!50!white}\textbf{4.62} \\
          $f_G$  &  [0.5 - 0.95]  &  \cellcolor{red!50!white}\textbf{0.9} \\
    \bottomrule
  \end{tabular}
  \label{tab:designs}
\end{table}
\begin{figure}[h]
    \centering
    \includegraphics[width=0.98\linewidth]{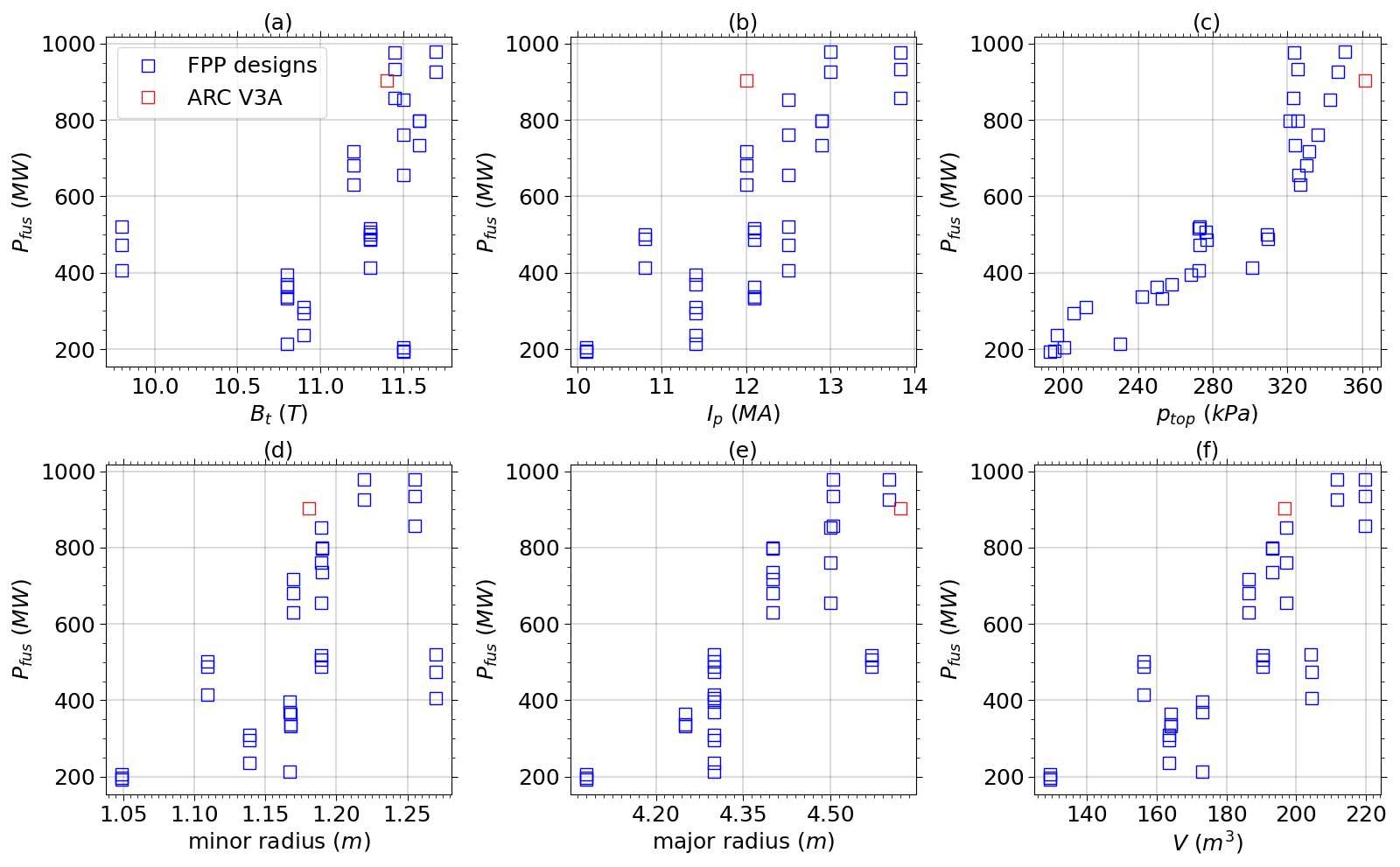}
    \caption{Fusion power vs: magnetic field (a), plasma current (b), pedestal top pressure (c), minor radius (d), major radius (e), plasma volume (f). The red point indicates a simulation inspired to ARC V3A, while the blue points refer to other ARC-class designs.}
    \label{fig:Pfus_dep}
\end{figure}
The resulting fusion power is shown in figure \ref{fig:Pfus_dep} for all the designs.
The red point in the figure indicates a design inspired by the ARC V3A, but with 5\% H concentration and the inclusion of impurity transport and detachment modeling, which leads to slightly different radial impurity profiles and separatrix conditions. In this figure, one can notice quite intuitive trends: the performance increases with magnetic field (a) and plasma current (b), as a combined effect of improved core-pedestal performance, resulting in higher pedestal top pressure (c); $P_{fus}$ also increases with minor radius (d) and major radius (e), since these conditions lead to higher volume (f). Since for most of simulations the pedestal is peeling-limited, higher elongations slightly improved the pedestal top pressure, while no trend was found for triangularity. In fact, for the explored designs, the triangularity has been observed to simply change the density at which the Peeling-Ballooning transition happens, without significantly impacting the maximum achievable pressure, at fixed elongation. The red point shows high performance, with $P_{fus}\sim 900\:MW$. Similar values were observed for ARC V3A in [Howard JPP 2026, accepted], where a simplified approach assuming fixed separatrix boundary conditions and impurity concentrations was adopted. However, it is worth mentioning that in [Howard JPP 2026, accepted] a wide variation of $P_{fus}$ was observed, spanning values from 600 to 1100 MW for different combinations of TGLF saturation rule and numerical settings, highlighting uncertainties in core transport predictions.

These results show different fusion targets for different high-field FPP configurations, including high-performing designs. However, additional uncertainties on physics inputs used in the simulations can affect $P_{fus}$, e.g. by modifying the pedestal stability. This highlights the importance of sensitivity studies on the physics parameters affecting the core-edge modeling, like $n_{sep}$, which will be explored in the next sections.

\section{Impact of core-edge modeling assumptions on performance}
\label{sec:results}
The simulations performed in the previous section assume a few inputs, whose quantification remains challenging with present models. These inputs have an impact on the predictions, since they affect the required SOL seeding, core impurity penetration / accumulation, radiation, pedestal stability and potentially the fusion performance. Therefore, in this section, a series of sensitivity studies on the input parameters is performed around the ARC V3A-like design point, to assess the variability of fusion power, H-mode access and detachment conditions.

\subsection{Scan of pedestal density}
\begin{figure}[h]
    \centering
    \includegraphics[width=0.98\linewidth]{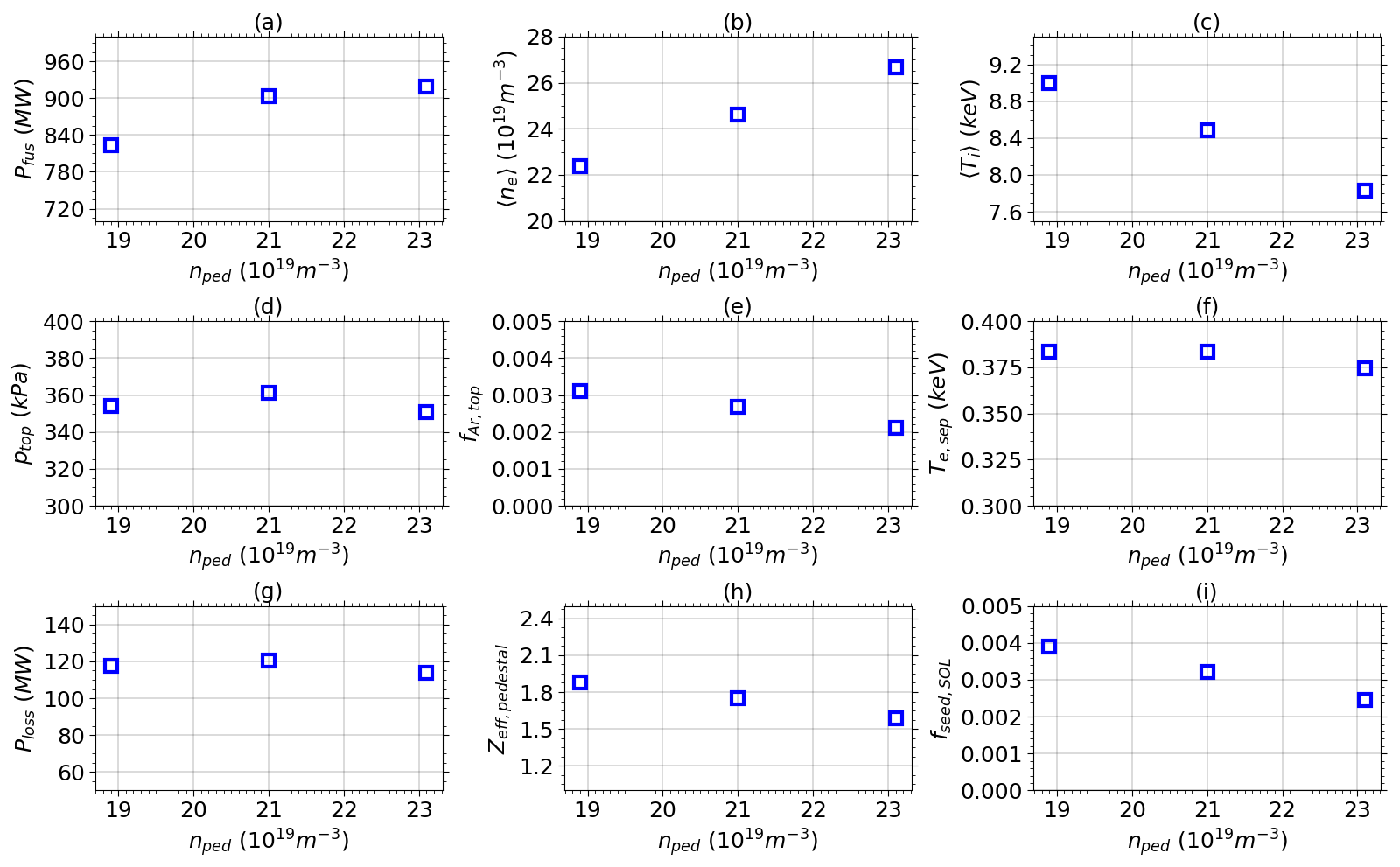}
    \caption{Global parameters for a scan of pedestal density: fusion power (a); volume average electron density (b); volume average ion temperature (c); top of pedestal pressure (d); concentration of seeded impurity at pedestal top (e); temperature at the separatrix (f); power at the separatrix (i.e. $P_{SOL}$ or $P_{loss}$) (g); $Z_{eff}$ at pedestal (h); seeded impurity concentration in the SOL (i).}
    \label{fig:V3A_SOL}
\end{figure}

\begin{figure}[h]
    \centering
    \includegraphics[width=0.48\linewidth]{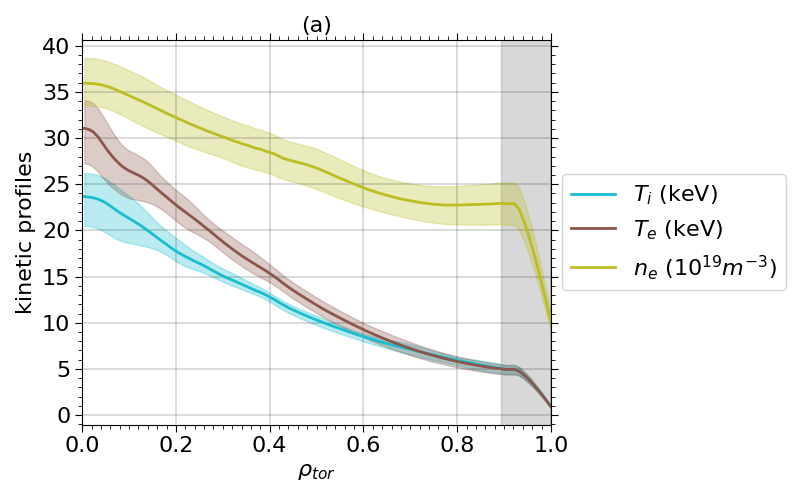}
    \includegraphics[width=0.48\linewidth]{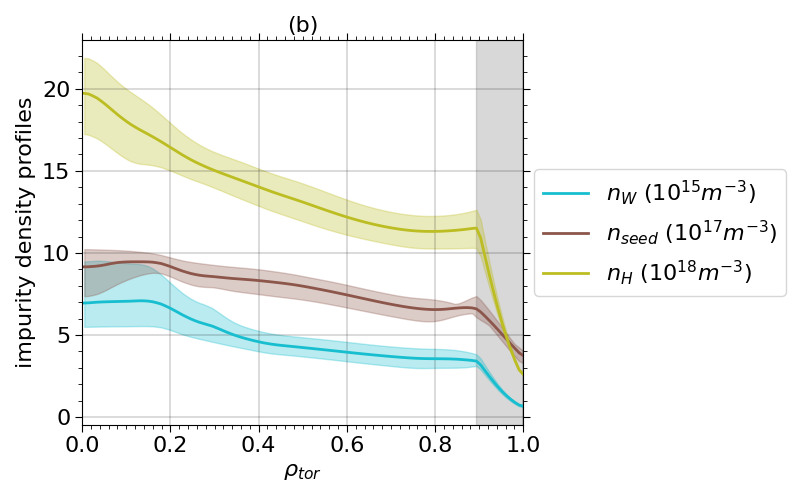}
    \includegraphics[width=0.48\linewidth]{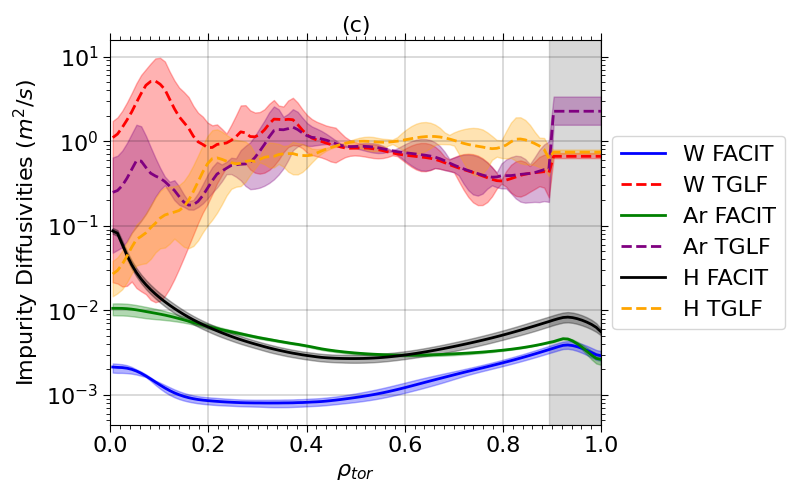}
    \includegraphics[width=0.48\linewidth]{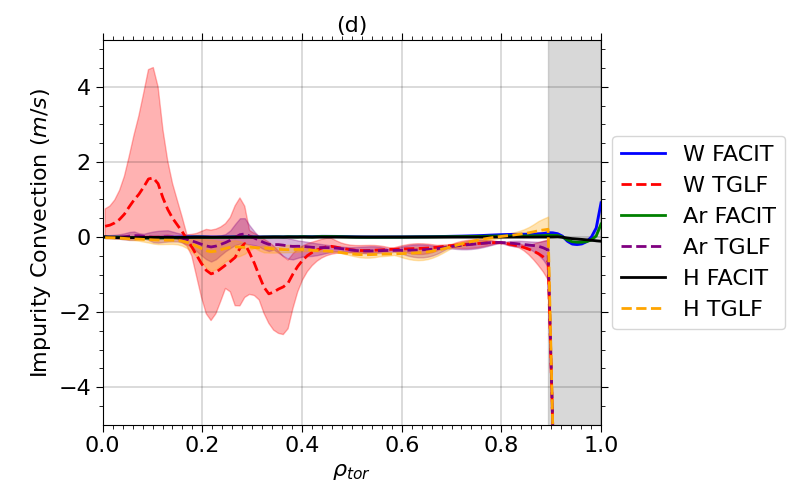}
    \caption{Profiles of: electron density, ion temperature and electron temperature (a); W, Ar and H density (b), neoclassical and turbulent impurity diffusivity for all species (c); neoclassical and turbulent impurity convection for all species (d). The lines indicate average profiles, while the shaded areas show the minimum and maximum values across the pedestal density scan. The gray vertical shade highlights the pedestal region. $\rho_{tor}$ is the square root of the normalized toroidal flux.}
    \label{fig:profiles_nscan}
\end{figure}

A scan in pedestal density (at fixed $n_{sep}/n_{ped}=0.4$) has been performed, using Ar as seeded species. $n_{ped}$ has been varied between $19\cdot10^{19}$ and $23\cdot10^{19}m^{-3}$, roughly resulting in $f_G=0.82$ and $0.97$. The results are shown in figure \ref{fig:V3A_SOL}.
An increase of fusion power (a) from $n_{ped}=19\cdot10^{19}$ to $21\cdot10^{19}m^{-3}$ is visible, while for higher density values, $P_{fus}$ is roughly constant, due to the transition to ballooning limited pedestals. In fact, the pedestal pressure (d) exhibits a maximum value for $n_{ped}=21\cdot10^{19}m^{-3}$ and a minimal variation across the scan. Since  $p_{top}$ shows very small variations, the higher average densities (b) are accompanied by lower average ion temperature (c). A very small impact has been found on the evolution of the SOL parameters. In particular, increasing $n_{ped}$ (at fixed $n_{sep}/n_{ped}$) leads to higher $n_{sep}$, with consequent lower $f_{Ar,SOL}$ needed for detachment (i), and lower $f_{Ar,top}$ (e). This, coupled with higher densities, provides a roughly constant radiation, causing a minimal deviation of $P_{loss}$ (g) and consequently similar values of separatrix temperature (f). \\
The kinetic profiles, together with the densities, diffusivities and convections of W, Ar and H are shown in figure \ref{fig:profiles_nscan}. Here, the lines indicate the average profiles, while the colored areas represent the minimum and maximum values across the density scan. Small deviations of the impurity profiles are found, exhibiting a moderate W density peaking, which corresponds to a near-axis concentration between $1.5\cdot10^{-5}$ and $2.5\cdot10^{-5}$ and is consistent with the pinch at $\rho_t=0.35$ in figure \ref{fig:profiles_nscan} (d). $f_H$ is roughly $0.05$ on the axis, which shows that a sufficient minority content can be achieved for efficient ICRH heating in the core, assuming $f_H=0.05$ at the top of the pedestal. Ar reaches a maximum concentration of $0.003$. Similar concentrations between axis and pedestal top have been found for all the impurities, showing that a simplified approach of assuming constant radial impurity concentrations can be used in future work to speed up simulations and derive large databases. A similar conclusion has been found for SPARC in \cite{muraca_impurity_2026}. Figure \ref{fig:profiles_nscan} (c) and (d) show that anomalous transport prevails on the neoclassical contribution in the core, which aligns with recent findings for ITER \cite{fajardo_theorybased_2025} and SPARC \cite{muraca_impurity_2026}. \\
The density scan has been performed up to $f_G=0.97$, to provide a safe margin with respect to the H-mode density limit, as done in [Hillesheim JPP 2026, accepted], [Howard JPP 2026, accepted]. However, $f_G>1$ has been reached in \cite{sauter_operation_2025}, and \cite{maris_realtime_2026} has shown that a different scaling based on edge collisionality can be used. Such finding has motivated additional simulations at $f_G>1$ to study the variation of fusion power at higher densities. Simulations from $f_G=0.97$ to $1.05$ show roughly the same $P_{fus}$. The reason behind this finding is that the scan of $n_{ped}$,  with fixed $n_{sep}/n_{ped}$ and roughly constant $Z_{eff}$ exhibits a transition to ballooning-limited stability. However, different selections of shaping and separatrix density could potentially retrieve higher fusion power at $f_G>1$, due to a higher top of pedestal pressure, if it remains peeling-limited. \\
This section has shown that small variations in pedestal density minimally affect the predicted fusion power and detachment conditions. However, two aspects of this initial modeling require a deeper investigation:
\begin{itemize}
    \item $n_{sep}/n_{ped}=0.4$ has been assumed across this scan. Nevertheless, this parameter strongly affects the pedestal pressure and the SOL seeding concentrations required to achieve detachment.
    \item The $f_{Ar,pedestal}$ values found in the simulations are high compared with existing literature from Alcator C-mod \cite{rice_very_2021}. Such high values are due to the enrichment factors computed with Eq. \ref{eq:Kallenbach_reg}, which show $\epsilon \sim 1.25$, while $\langle \epsilon \rangle \sim 3$ was found for AUG in \cite{kallenbach_divertor_2024}. This discrepancy is related to the inverse dependence of enrichment on the divertor neutral pressure (see Eq. \ref{eq:Kallenbach_reg}), which leads to lower $\epsilon$ in ARC.
\end{itemize}
These observations motivate additional scans in $n_{sep}/n_{ped}$ (with fixed $n_{ped}$) and SOL-core enrichment factor, which will be shown in the next sections.

\subsection{Scan of separatrix density}
\begin{figure}[h]
    \centering
    \includegraphics[width=0.98\linewidth]{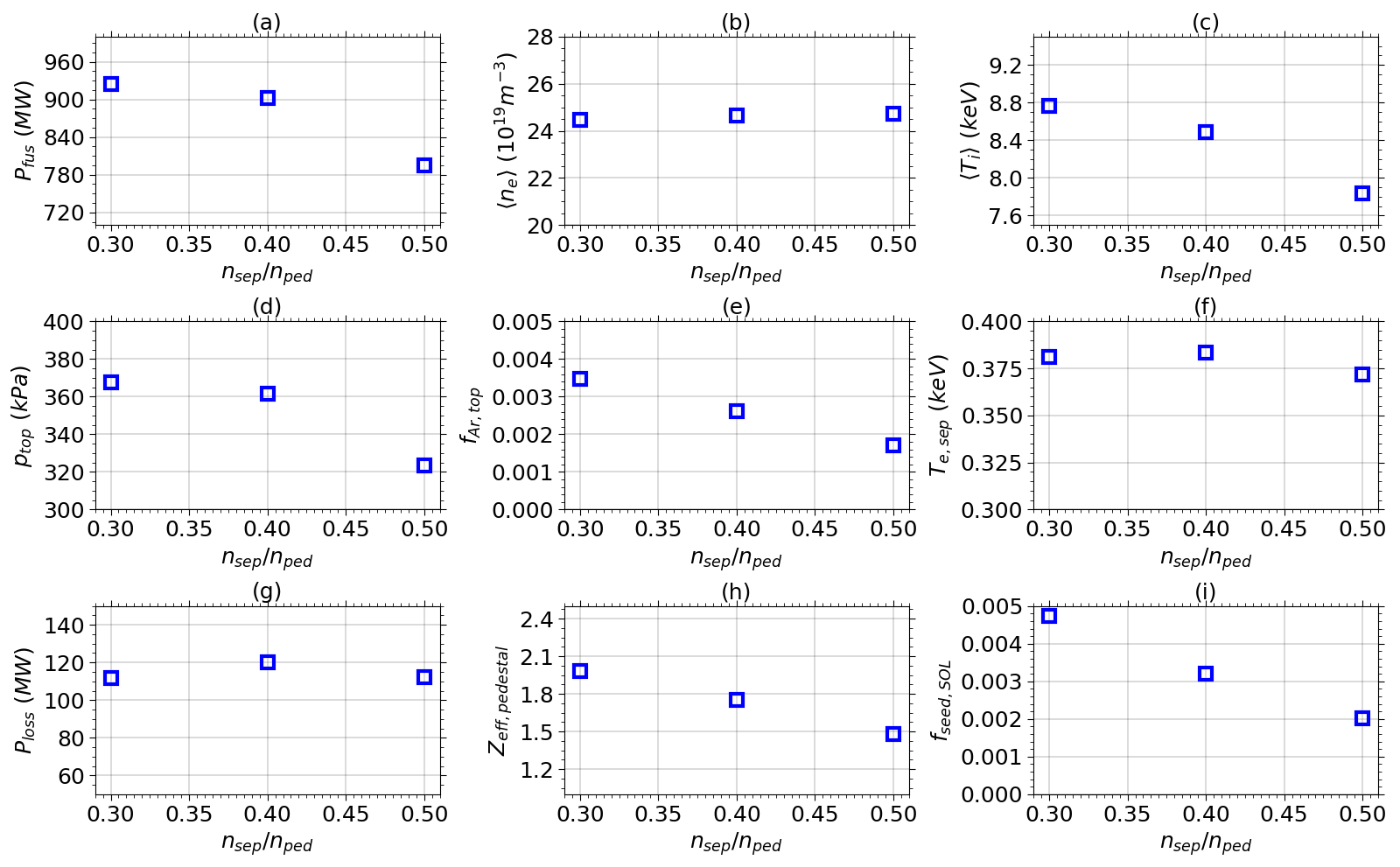}
    \caption{Global parameters for a scan of separatrix density: fusion power (a); volume average electron density (b); volume average ion temperature (c); top of pedestal pressure (d); concentration of seeded impurity at pedestal top (e); temperature at the separatrix (f); power at the separatrix (i.e. $P_{SOL}$ or $P_{loss}$) (g); $Z_{eff}$ at pedestal (h); seeded impurity concentration in the SOL (i).}
    \label{fig:V3A_nsep}
\end{figure}

\begin{figure}[h]
    \centering
    \includegraphics[width=0.49\linewidth]{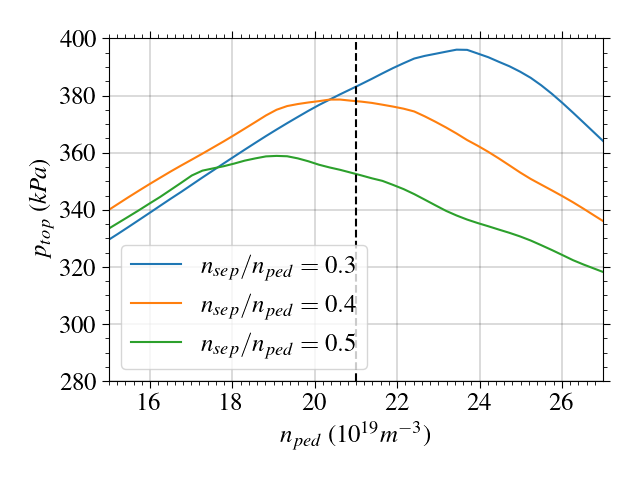}
    \includegraphics[width=0.49\linewidth]{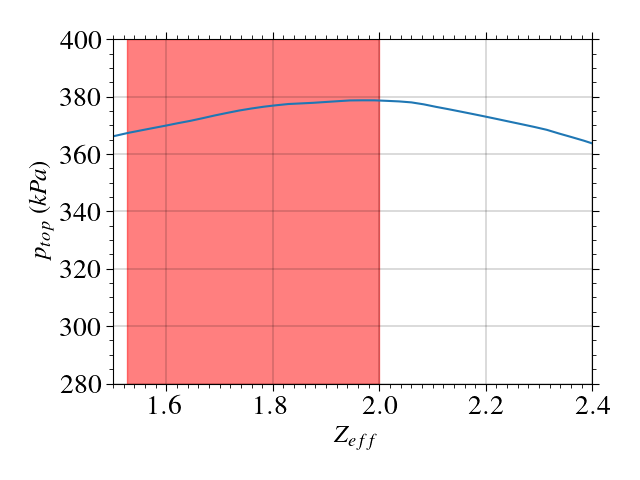}
    \caption{Left: top of pedestal pressure vs pedestal density for different $n_{sep}/n_{ped}$ ratios; right: $p_{top}$ vs $Z_{eff}$ for $n_{sep}/n_{ped}=0.4$ and $n_{ped}=21\cdot 10^{19}m^{-3}$. }
    \label{fig:ptop_vs_nped}
\end{figure}

A scan in $n_{sep}$, keeping fixed pedestal density ($n_{ped}=21\cdot10^{19}m^{-3}$, $f_G=0.9$), has been performed to assess its effect on pedestal top pressure, fusion performance and exhaust handling. To select a range for the scan, $n_{sep}$ has been computed with Eq. 8 from \cite{kallenbach_neutral_2019} and Eq. 6 from \cite{silvagni_predictive_2026}, using geometrical and physics inputs consistent with [Eich JPP 2026, accepted]. Employing both formulae, and assuming $P_{loss}=\left[60 - 180\right]\:MW$ and $p_{0,div}=\left[10 - 30\right]\:Pa$, an $n_{sep}$ spanning between $4\cdot 10^{19}$ and $12\cdot 10^{19}m^{-3}$ has been found. Therefore, a scan in $n_{sep}$ has been performed, assuming a conservative range between 6.3 and 10.5 (i.e. $n_{sep}/n_{ped}=0.3$ and $0.5$). The results are shown in figure \ref{fig:V3A_nsep}.
While similar values are found for $n_{sep}/n_{ped}\leq0.4$, the simulation with $n_{sep}/n_{ped}=0.5$ shows a clear drop in $p_{top}$ (d) and $P_{fus}$ (a) to 330 kPa and 800 MW respectively. This change is due to the modification of the peeling-ballooning stability curve at different $n_{sep}/n_{ped}$ values, that can be seen in the left plot of figure \ref{fig:ptop_vs_nped}. This figure shows that similar pressure can be reached at $n_{ped}=21\cdot10^{19}m^{-3}$ for $n_{sep}/n_{ped}=[0.3-0.4]$, while for $n_{sep}/n_{ped}=0.5$ the pedestal is ballooning-limited and the pressure is lower. \\
The drop in $P_{fus}$ is also reflected in the volume averaged ion temperature (Figure \ref{fig:V3A_nsep} c). An increased separatrix density requires lower concentration of the seeded impurity (i) to radiate in the SOL and access detachment. This reduction is reflected in the lower top of pedestal concentrations (e) and $Z_{eff,pedestal}$ (h), that affects the pedestal prediction. However, differences in $Z_{eff}$ in these conditions have a small impact, as can be seen in the right plot of figure \ref{fig:ptop_vs_nped}. Here, the red colored region covers the $Z_{eff}$ values spanned across the $n_{sep}$ scan, indicating a small variation of $p_{top}$ ($\sim15\:kPa$) with fixed $n_{sep}/n_{ped}=0.4$. \\
The lower impurity concentration in the core radiates less power, compensating the lower $P_{fus}$ and leading to similar values of $P_{loss}$, as depicted in figure \ref{fig:V3A_nsep} (g). Since $P_{loss}$ is roughly constant, the separatrix temperature does not change sensitively (f). \\
In conclusion, although higher $n_{sep}/n_{ped}$ values can lower performance, high fusion power (i.e. $ \sim 800MW$) can still be reached with Ar-seeded plasmas, while applying less strict constraints on the seeding level needed to reach detachment.

\subsection{Scan of enrichment factor}
\begin{figure}[h]
    \centering
    \includegraphics[width=0.98\linewidth]{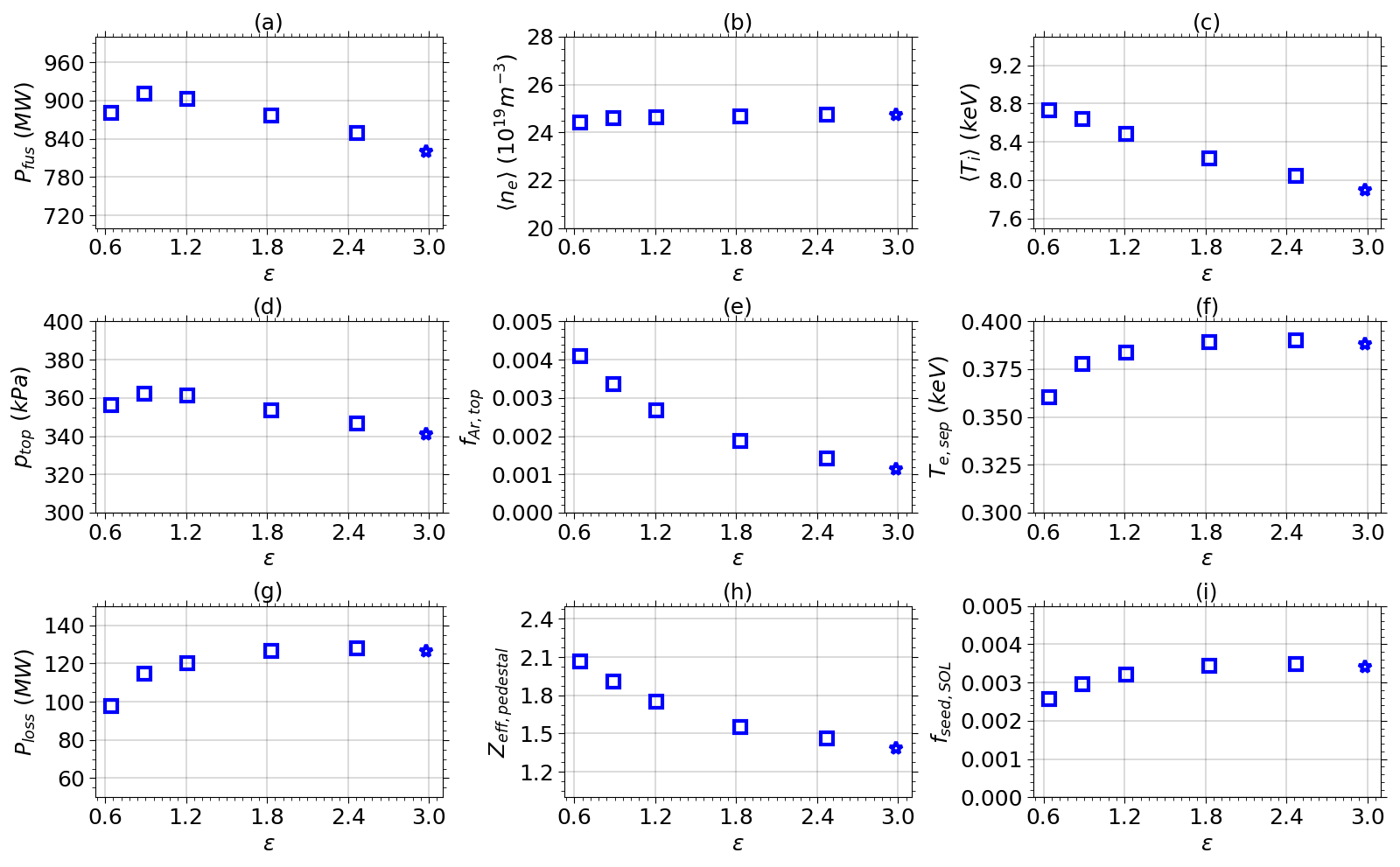}
    \caption{Global parameters for a scan of enrichment factor: fusion power (a); volume average electron density (b); volume average ion temperature (c); top of pedestal pressure (d); concentration of seeded impurity at pedestal top (e); temperature at the separatrix (f); power at the separatrix (i.e. $P_{SOL}$ or $P_{loss}$) (g); $Z_{eff}$ at pedestal (h); seeded impurity concentration in the SOL (i). The star indicates the average enrichment value from figure 8 in \cite{kallenbach_divertor_2024}.}
    \label{fig:V3A_fcore}
\end{figure}

The exact determination of the enrichment factor ($\epsilon$) is a challenging task, that depends on impurity penetration and seeding rate, pedestal formation and physics, and transport mechanisms which require time-dependent modeling of the pulse. Since such modeling is beyond the scope and capability of the present framework, a wide range of enrichment factors has been explored, to address uncertainties and their effect on fusion performance.
A scan of $\epsilon$ between 50 and 200\% of the value calculated with Eq. \ref{eq:Kallenbach_reg} has been performed. It is worth mentioning that the lowest value included is roughly consistent with that computed using Eq. 10 of \cite{kallenbach_divertor_2024}, which comes from a balance model that includes simple expressions for the SOL particle pinch and the core impurity confinement time \cite{rice_core_2015}. The low enrichment found with this formula comes from the higher core impurity confinement time found for ARC with respect to AUG. Additionally to this scan, a simulation with $\epsilon=3$ has been included, to reproduce values of enrichment qualitatively consistent with those found experimentally in AUG, when using Ar seeding. \\
The results of the scan, with fixed $n_{ped}=21\cdot10^{19}m^{-3}$ and $n_{sep}/n_{ped}=0.4$, are shown in figure \ref{fig:V3A_fcore}.
The deviation of fusion power (a) is surprisingly small, varying between 820 and 920 MW. This variation comes from corresponding different values of pedestal top pressure (d), which are in turn related to different $Z_{eff,pedestal}$ (h). This effect is reflected also in the average ion temperature (c). Increasing the enrichment, an increase in $P_{loss}$ is found (g), due to a lower core penetration (e) and radiation, with a subsequent increase of separatrix temperature (f). Since the separatrix density is fixed, the SOL seeding concentration is roughly constant (i), except for very low enrichment and $P_{loss}$ values. \\
In synthesis, variations of Ar enrichment have a small impact on the overall performance, showcasing robust core-edge integrated solutions that lead to high fusion power ($820<P_{fus}<920\:MW$), while maintaining detached conditions with sufficient impurity seeding.

\subsection{Assumption of Ne-seeded plasmas}
The simulations shown in the previous sections assume Ar as seeded impurity. However, similar calculations should be done to assess the feasibility of a high performing detached scenario with different species. Therefore, in this section, plasmas with Ar- and Ne-seeding are compared. An initial set of runs with $n_{sep}/n_{ped}=0.4$ and $n_{ped}=21\cdot10^{19}m^{-3}$ have been executed, using 5 different enrichment values, corresponding to 50, 75, 100, 150, 200\% of the value computed by Eq. \ref{eq:Kallenbach_reg} within the simulation. Two additional runs with fixed $\epsilon=3$ and $1$ have been performed, to reproduce the enrichment values found experimentally in AUG with Ar- and Ne-seeding respectively \cite{kallenbach_divertor_2024}. For each of these simulations, additional scans of $n_{sep}/n_{ped}$ and $n_{ped}$ have been conducted separately, leading to 30 cases for each species, i.e. 6 enrichment variations $\times$ (1 nominal case $+$ 2 $n_{sep}/n_{ped}$ values $+$ 2 $n_{ped}$ values)).
\begin{figure}[h]
    \centering
    \includegraphics[width=0.98\linewidth]{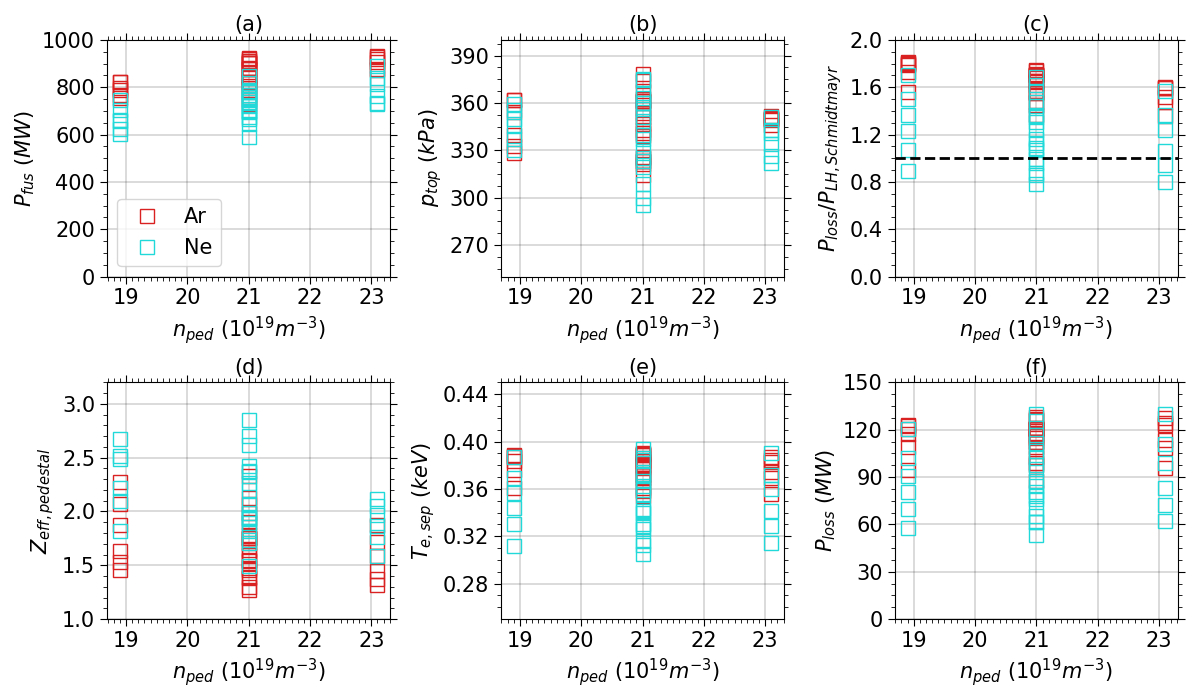}
    \caption{Fusion power (a), top of pedestal pressure (b), $P_{loss}/P_{LH,Schmidtmayr}$ (c), $Z_{eff}$ at pedestal (d), separatrix temperature (e), power loss (f) as function of pedestal density. Different colors indicate different seeded species.}
    \label{fig:global_fG}
\end{figure}
The results are summarized and compared for the two seeded species in figures \ref{fig:global_fG}, \ref{fig:global_epsilon} and \ref{fig:global_nsep}. These figures illustrate the trends of the principal global plasma parameters as functions of $n_{ped}$, $\epsilon$ and $n_{sep}/n_{ped}$, respectively. Note that the figures include all the simulations performed varying the three mentioned input parameters, explaining why a high variability of global parameters is found at fixed $n_{ped}$, $\epsilon$ and $n_{sep}/n_{ped}$. \\
\begin{figure}[h]
    \centering
    \includegraphics[width=0.98\linewidth]{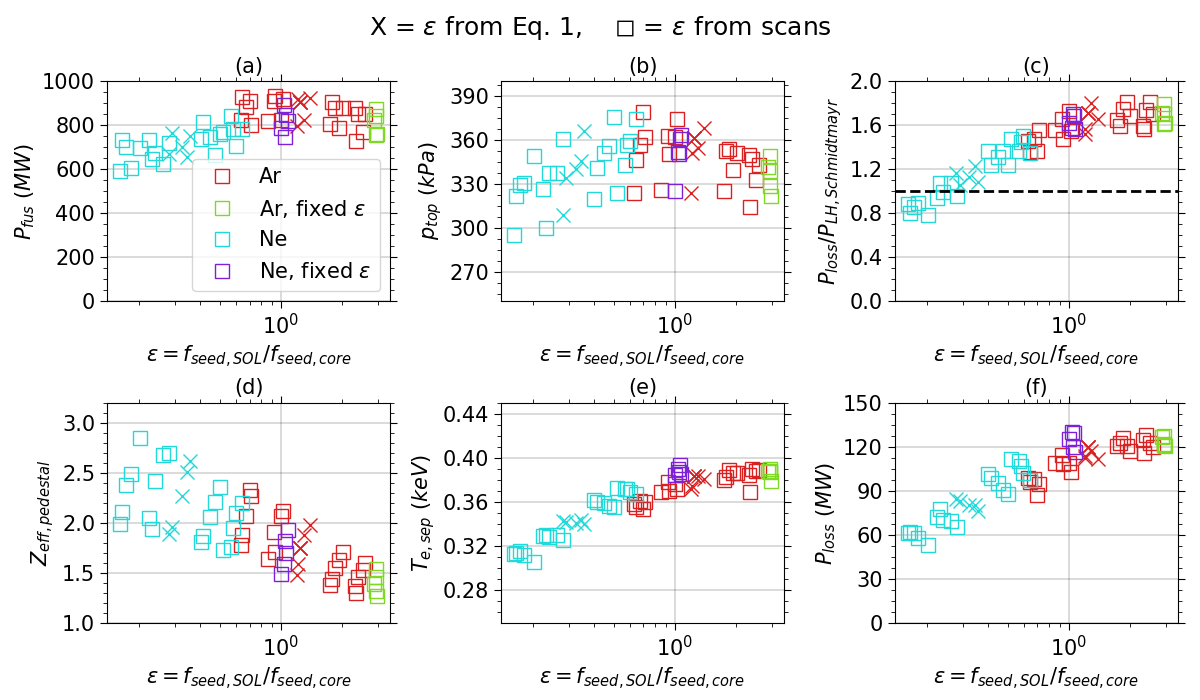}
    \caption{Fusion power (a), top of pedestal pressure (b), $P_{loss}/P_{LH,Schmidtmayr}$ (c), $Z_{eff}$ at pedestal (d), separatrix temperature (e), power loss (f) as function of $\epsilon$. Red and blue squares indicate scans of $\epsilon$ for Ar- and Ne-seeded plasmas. Green and purple squares are simulations with $\epsilon$ equal to 3 (Ar seeding) and 1 (Ne seeding) respectively. The crosses indicate simulations where $\epsilon$ is computed with Eq. \ref{eq:Kallenbach_reg}.}
    \label{fig:global_epsilon}
\end{figure}
\begin{figure}[h]
    \centering
    \includegraphics[width=0.98\linewidth]{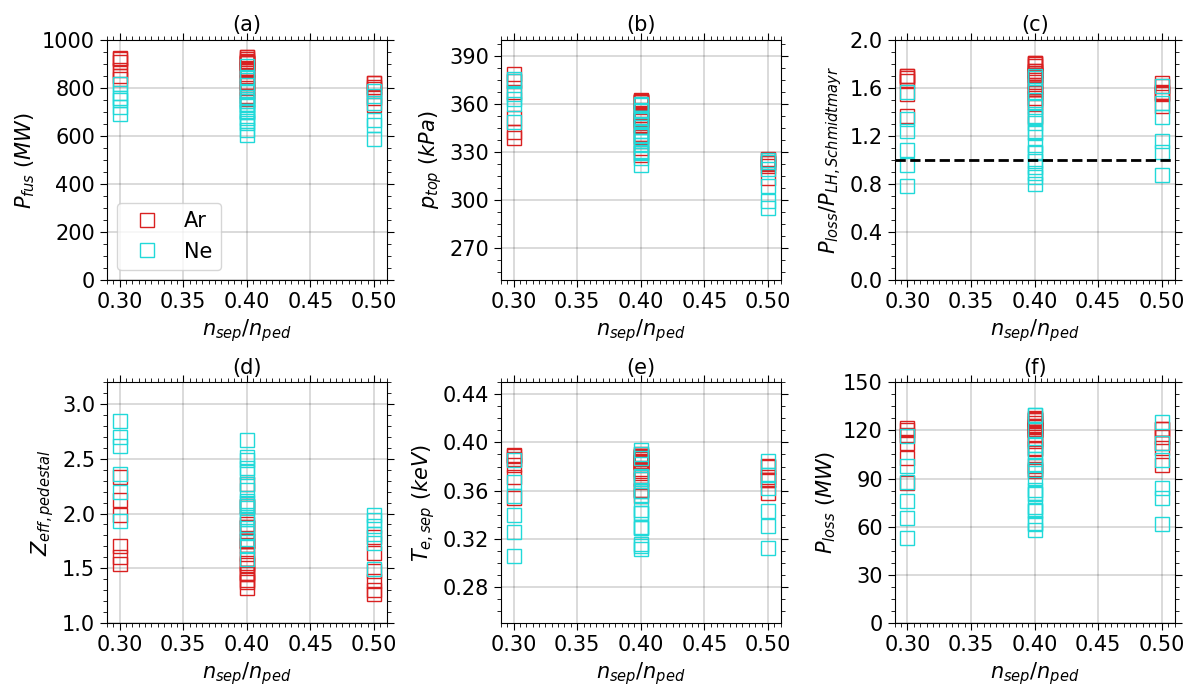}
    \caption{Fusion power (a), top of pedestal pressure (b), $P_{loss}/P_{LH,Schmidtmayr}$ (c), $Z_{eff}$ at pedestal (d), separatrix temperature (e), power loss (f) as function of separatrix density. Different colors indicate different seeded species.}
    \label{fig:global_nsep}
\end{figure}
Figure \ref{fig:global_fG} shows a slightly higher $P_{fus}$ (a) increasing $n_{ped}$ for both species, until $n_{ped}=21\cdot10^{19}m^{-3}$, which is near the transition from peeling to ballooning limited pedestals, as shown by the roughly constant $p_{top}$ at different $n_{ped}$ values in (b). The highest performance is reached using Argon, while lower $P_{fus}$ is obtained with Ne. The main reason for this difference is the lower DT concentrations in Ne-seeded cases, which will be discussed later in detail. \\
Surprisingly, a higher $Z_{eff,pedestal}$ is found for Ne (d), even though its atomic number is lower than Ar. This finding is justified by the Ne lower enrichment factors, as can be seen in figure \ref{fig:global_epsilon}. In this figure, $\epsilon$ orders the results for the two species. Although 6 enrichment variations have been mentioned, more $\epsilon$ values are present in the figure. In fact, in each simulation a "nominal value" is initially computed with Eq. \ref{eq:Kallenbach_reg} and then a scaling factor spanning from 50 to 200\% is applied to obtain $\epsilon$, but the nominal value is affected by different divertor neutral pressure values computed by the X-Lengyel model. Moreover, additional small numerical deviations result from the feedback algorithm. This is particularly evident in figure \ref{fig:global_epsilon} (d) and (f), where small groups of points surround average $\epsilon$ values, but deviate slightly because of different $n_{sep}/n_{ped}$ and $n_{ped}$ assumptions, which impact X-Lengyel calculations. The low enrichment values found for Ne reflect a high core accumulation with higher concentrations in the core than in the edge. It is worth mentioning that these levels of accumulations are rarely observed in present devices. The lower enrichment values of Ne-seeded plasmas lead to higher $Z_{eff,pedestal}$ (d), lower minimum pedestal (b), lower fusion (a), lower $P_{loss}$ (f), lower $T_{sep}$ (e) and lower $f_{LH,Schmidtmayr}$ (c), where $f_{LH,Schmidtmayr}=P_{i,loss}/P_{LH,Schmidtmayr}$, $P_{i,loss}$ is the ion power loss entering the SOL and $P_{LH,Schmidtmayr}$ is the LH power threshold according to the Schmidtmayr scaling \cite{schmidtmayr_investigation_2018}. \\
Figure \ref{fig:global_nsep} shows expected trends for both seeding species: $p_{top}$ (b) and  $P_{fus}$ (a) decrease with $n_{sep}/n_{ped}$, as previously observed. Moreover, a small decrease of $Z_{eff,pedestal}$ (d) is found increasing the separatrix density, due to the lower impurity seeding needed to detach. \\
As shown in (a) and (c) plots of figures \ref{fig:global_fG}, \ref{fig:global_epsilon} and \ref{fig:global_nsep}, Ar-seeded plasmas experience higher fusion power and allow robust H-mode access, while with Ne seeding $P_{fus}$ is frequently below $800\:MW$ and the plasma is marginally accessing H-mode ($f_{LH,Schmidtmayr}\geq1$). Nevertheless, conservative values of the enrichment have been adopted across most of the simulations, while assuming $\epsilon=1$ for Ne-seeded cases, consistently with AUG results \cite{kallenbach_divertor_2024}, leads to $f_{LH,Schmidtmayr}\sim1.6$, as shown by the purple squares in figure \ref{fig:global_epsilon} (c). Using the Delabie scaling for the LH transition power \cite{delabie_empirical_2026} leads to average $f_{LH}$ values of 1.84 and 1.6 for Ar- and Ne-seeded plasmas. Using another Delabie threshold that keeps into account radiation losses leads to more pessimistic results, with $\langle f_{LH}\rangle$ equal to 1 and 0.7 for Ar and Ne seeding. However, this threshold shows higher mean root square error compared to the expression without radiation. The Martin scaling \cite{martin_power_2008} gives similar results to Delabie, with an average $f_{LH}$ equal to 1 for Ar- and 0.74 for Ne-seeded plasmas. Nevertheless, the Martin scaling has wide error bars and is not fitted on W-wall machines. These estimates suggest that there are uncertainties related to H-mode access, and future investigations with time-dependent modeling are required to precisely determine the LH transition. In general, a higher available ICRH power could be beneficial to address these uncertainties. \\
In order to investigate further confinement, performance and H-mode access, additional global parameters are summarized in table \ref{tab:globals_species} and the ion temperature and electron density profiles are shown in figure \ref{fig:profiles_Ar_Ne}.
\begin{table}[h]
  \centering
  \caption{Average values and standard deviations of core global parameters and peaking factors for Ar- and Ne-seeded plasmas, across scans in pedestal density, impurity enrichment and separatrix density. The variation of the kinetic profiles is shown in figure \ref{fig:profiles_Ar_Ne}.}
  \begin{tabular}{>{\columncolor{lightblue}}c | c | c | c | c | c}
    \toprule
    \rowcolor{lightblue} 
    \textbf{Parameter} & \textbf{$\langle ...\rangle_{Ar}$} & \textbf{$\langle ...\rangle_{Ne}$} & \textbf{$\sigma_{Ar}$} & \textbf{$\sigma_{Ne}$} \\
    \midrule
          $\langle n_e\rangle$ ($10^{19}m^{-3}$) & 24.4 & 24.1 & 0.26 & 0.14 \\
          $\langle T_i\rangle$ ($keV$) & 8.6 & 9.3 & 0.57 & 0.57 \\
          $\nu_{n_e}$ & 1.34 & 1.29 & 0.04 & 0.01 \\
          $\nu_{T_i}$ & 2.05 & 2.11 & 0.04 & 0.04 \\
          $\langle f_{DT}\rangle$ & 0.89 & 0.77 & 0.026 & 0.047 \\
          $\langle Z_{eff}\rangle$ & 2 & 2.6 & 0.38 & 0.43 \\
          $P_{rad}$ ($MW$) & 104 & 105 & 18 & 15 \\
    \bottomrule
  \end{tabular}
  \label{tab:globals_species}
\end{table}
\begin{figure}[h]
    \centering
    \includegraphics[width=0.48\linewidth]{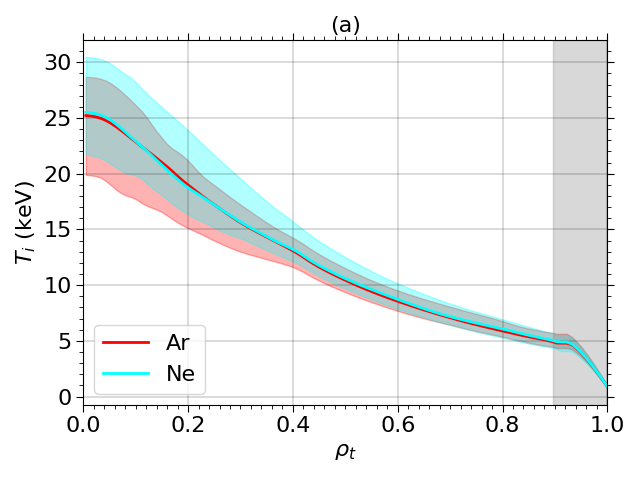}
    \includegraphics[width=0.48\linewidth]{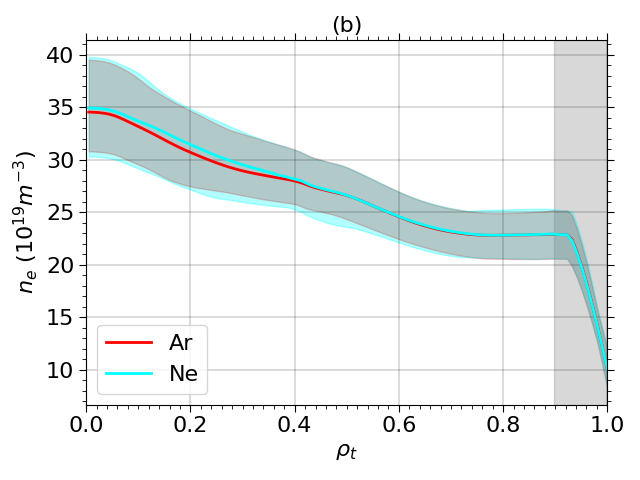}
    \caption{Ion temperature (a) and electron density (b) profiles across $n_{ped}$, $n_{sep}$ and $\epsilon$ scans for the Ar-seeded (red) and Ne-seeded (cyan) plasmas. The solid lines are the nominal profiles, while the shaded area indicates the maximum and minimum values. The grey colored area indicates the pedestal region. $\rho_t$ is the square root of the normalized toroidal flux.}
    \label{fig:profiles_Ar_Ne}
\end{figure}
In table \ref{tab:globals_species}, slightly different values of volume averaged $n_e$ and electron density peaking ($\nu_{n_e}$) are found for Ar- and Ne-seeded plasmas, due to differences in $Z_{eff}$ and DT dilution. Higher $T_i$ and ion temperature peaking ($\nu_{T_i}$) with Ne seeding are related to higher $Z_{eff}$ and fuel dilution (i.e. $f_{DT}$) in the core, which lead to a different turbulent regime \cite{jenko_critical_2001, fable_role_2010, ennever_effects_2015, kim_fullf_2017, rodriguez-fernandez_core_2024}. This can be noticed also in figure \ref{fig:profiles_Ar_Ne}, where the colored region around the nominal profiles, that indicates the variations across the scan, is higher for Ne-seeded plasmas. Interestingly, although Ne simulations show a higher volume average ion temperature, they still exhibit lower performance. This is related to the lower DT concentration, which provides better confinement \cite{ennever_effects_2015, kim_fullf_2017, rodriguez-fernandez_core_2024} but less fuel for fusion reactions, with the latter dominating on the former. As previously mentioned, this is the main reason for the lower fusion power found for Ne-seeded plasmas. However, higher values of DT concentration can be reached with lower Ne accumulation, together with higher top of pedestal pressure (due to lower $Z_{eff,pedestal}$). These two combined effects allow the plasma to retrieve higher fusion power conditions ($P_{fus}>800\:MW$), together with robust H-mode access ($f_{LH,Schmidtmayr}\sim1.6$). Simulations with Ar seeding show a lower average $Z_{eff}$ across all scans, providing less stringent solenoid flux consumption constraints for a long-pulse [Hillesheim JPP 2026, accepted], but similar values can be reached in highly enriched Ne-seeded plasmas, as suggested by the high standard deviation of $Z_{eff}$ in table \ref{tab:globals_species}. Finally, the radiation power is similar for Ne and Ar because the higher impurity accumulation of the former counterbalances the higher charge of the latter. Since Ne-seeded plasmas exhibit lower fusion, this results in the lower $P_{loss}$ and $f_{LH, Schmidtmayr}$ shown in figures \ref{fig:global_fG}, \ref{fig:global_epsilon}, \ref{fig:global_nsep} (f) and (c). \\
Although N is not considered for ARC operation, due to the difficulties connected with the extraction of tritiated ammonia in DT-fueled tokamak environments \cite{park_selfconsistent_2017}, additional simulations have been performed with a N-equivalent radiator, to verify how impurity core penetration compares with AUG \cite{kallenbach_divertor_2024} and ITER \cite{sytova_comparing_2019} findings. Using the same framework, simulations have been performed scanning $n_{sep}/n_{ped}$, $n_{ped}$, and $\epsilon$, showing trends similar to those discussed earlier, an average enrichment factor of 3, $P_{fus}$ spanning from 760 to 1000 MW and robust H-mode conditions, with $\langle P_{rad}\rangle=80\:MW$ and $f_{LH,Schmidtmayr}$ varying between 1.3 and 2. Similar kinetic profiles to Ar-seeded plasmas are found, with $f_{DT}=0.84$, $\langle Z_{eff}\rangle<2$ and slightly higher W and N density peaking.
Simulations with Kr have not been considered due to the excessive core radiation expected employing the enrichment factors computed as in \cite{kallenbach_divertor_2024}. \\
In summary, the findings described in this section highlight that, under the assumptions of this study, high fusion performance and robust H-mode access can be obtained with Ar seeding, achieving detachment and divertor protection, while Ne-seeded plasmas show lower performance, exhibiting $600<P_{fus}<850\:MW$ and $0.75<f_{LH,Schmidtmayr}<1.75$.

\subsection{Investigation of core impurity transport}
\begin{figure}[h]
    \centering
    \includegraphics[width=0.98\linewidth]{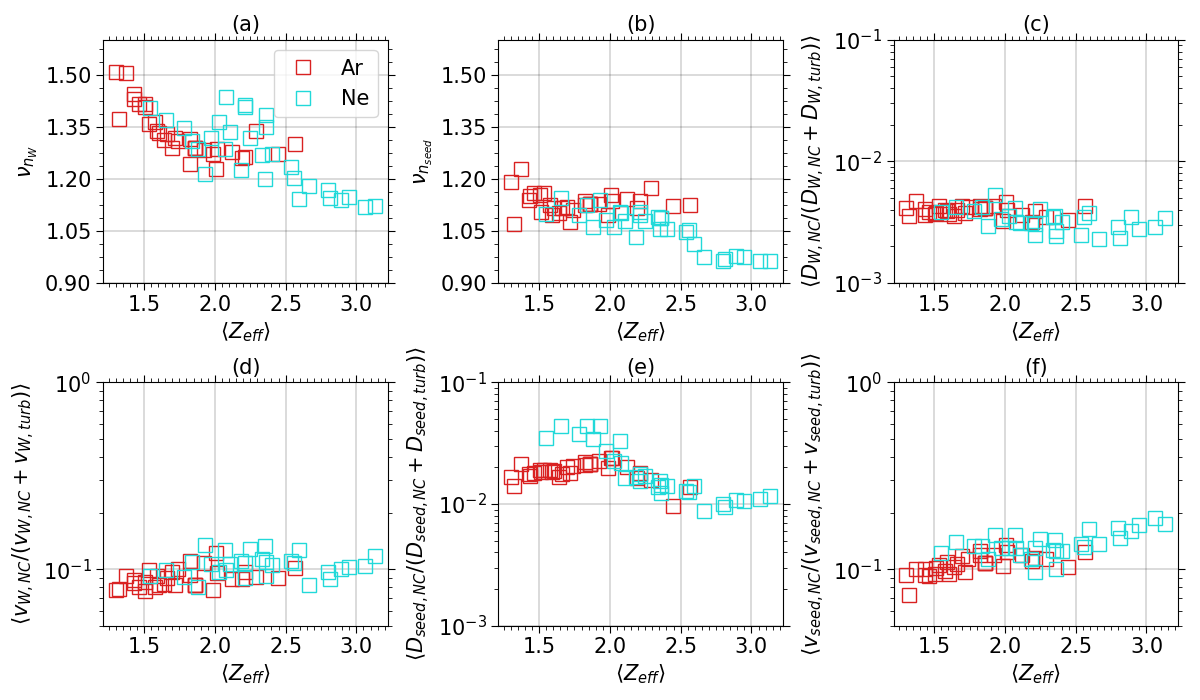}
    \caption{W (a) and seeded species (b) density peaking factors, neoclassical to total ratio of W diffusivity (c) and convection (d), neoclassical to total ratio of seeded species diffusivity (e) and convection (f), as function of volume average $Z_{eff}$. Different colors indicate different seeded species. (a)-(b), (c)-(e) and (d)-(f) are plotted on the same scales to emphasize their relative magnitude.}
    \label{fig:global_imps}
\end{figure}

In order to summarize the information from impurity transport in the core, the W and seeded-species density peaking, here defined as $n_{imp,0}/\langle n_{imp}\rangle$, where $n_{imp,0}$ is the density on-axis and "imp" refer to either W or Ar/Ne, are shown in figure \ref{fig:global_imps}, together with $\langle D_{NC}/(D_{NC}+D_{turb})\rangle$ and $\langle \left| v_{NC} \right| /(\left| v_{NC} \right|+\left| v_{turb} \right|)\rangle$. $\langle\rangle$ indicates a volume average from the magnetic axis to top of pedestal, $D_{NC}$ ($v_{NC}$) and $D_{turb}$ ($v_{turb}$) indicate the neoclassical and turbulent components of the diffusivity (convection). 
The W core impurity peaking shows some variation, predicting lower values at higher $Z_{eff}$ and with Ne seeding. This effect is expected to be related to a change in the balance of ITG and TEM. The neoclassical diffusivities are negligible compared to the turbulent ones for both W (c) and seeded species (e), as expected. The convection ratios show similar ranges for W and Ar/Ne, with values below 0.15 for the former (d) and 0.2 for the latter (f). A weak increase of the convection ratio is found with $Z_{eff}$, probably related to different values of density and temperature peaking factors (see table \ref{tab:globals_species}). \\
TGLF standalone simulations have been performed, scanning $Z_{eff}$ at $\rho_t=0.4$ for the Ar-seeded plasma with $n_{ped}=21\cdot10^{19}m^{-3}$ (i.e. $f_G=0.9$).
\begin{figure}[h]
    \centering
    \includegraphics[width=0.49\linewidth]{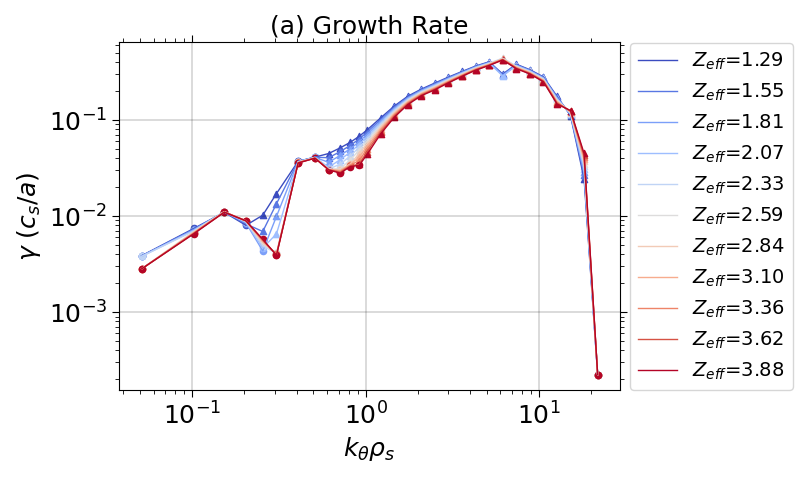}
    \includegraphics[width=0.49\linewidth]{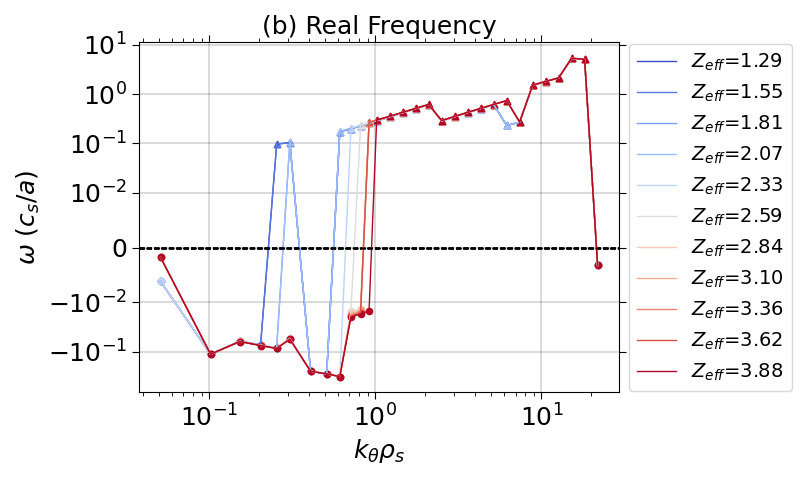}
    \includegraphics[width=0.49\linewidth]{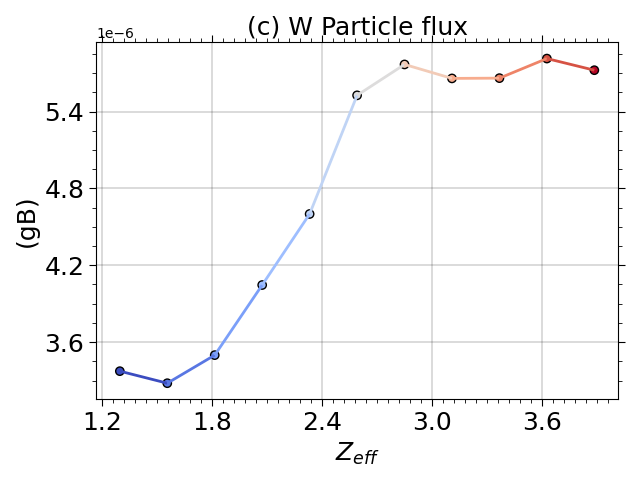}
    \includegraphics[width=0.49\linewidth]{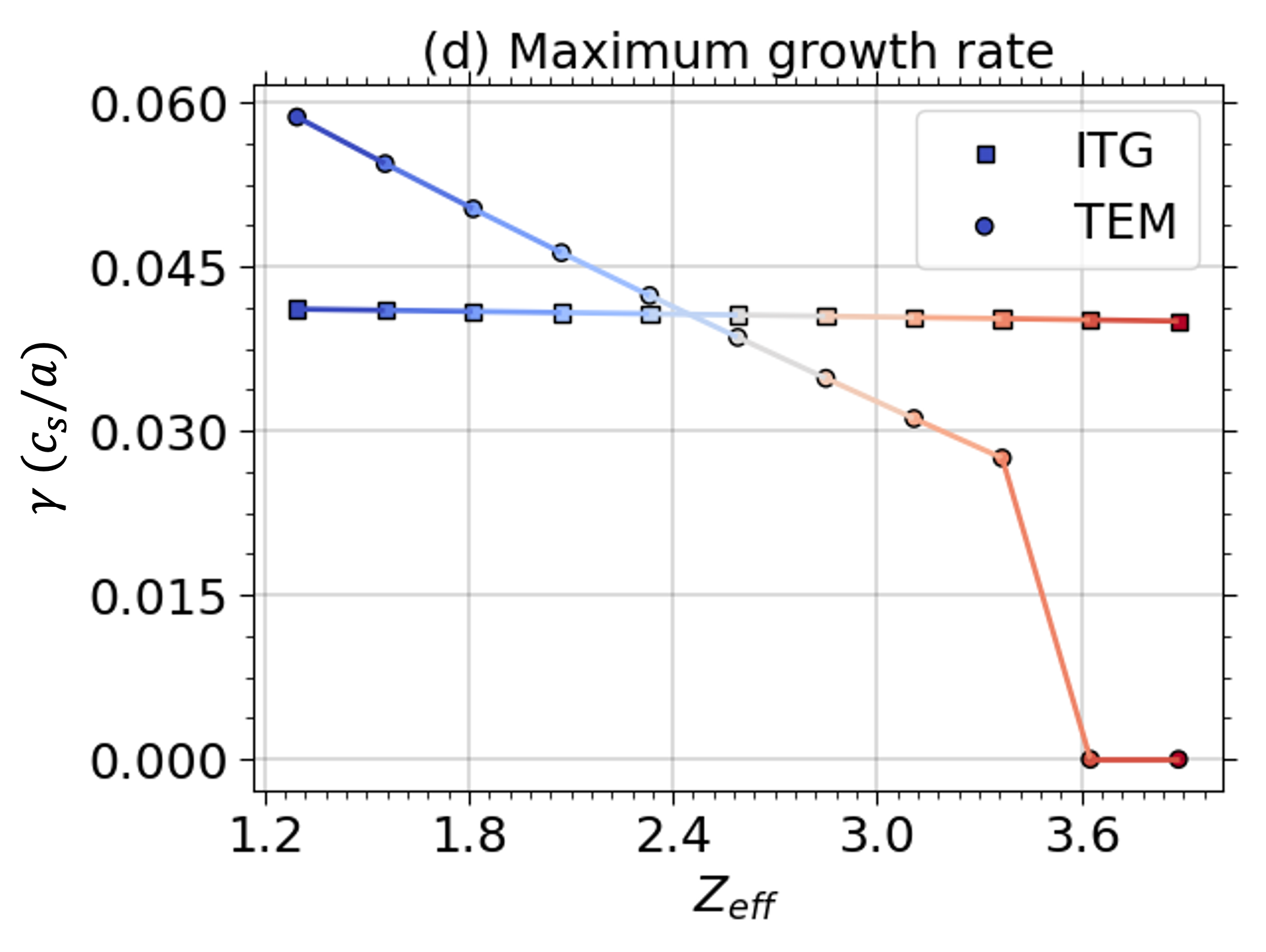}
    \caption{Results of TGLF standalone simulations, scanning $Z_{eff}$ at $\rho_t=0.4$, for an Ar-seeded plasma at $f_G=0.9$. (a) and (b) show the growth rate and frequency spectra, (c) depicts the W particle flux and (d) illustrates the most unstable growth rates for ITG and TEM instabilities.}
    \label{fig:TGLF_std}
\end{figure}
The results of the scan are shown in figure \ref{fig:TGLF_std}, where the growth rate (a) and mode frequency (b) spectra are depicted, together with the W particle flux (c) and the most unstable growth rates for ITG and TEM instabilities (d), which have been assumed to exist for $k_y\rho_s<1$. Increasing $Z_{eff}$, a higher W particle flux is found. Surprisingly, a transition to a TEM-dominated regime has been found for $Z_{eff}<2.4$, although the turbulence remains characterized by a mixed ITG/TEM regime.
The increase of W flux with $Z_{eff}$ is consistent with figure \ref{fig:global_imps} (a), which shows a slightly lower W peaking at higher effective charge. The variation of seeded-impurity peaking with $Z_{eff}$ is more challenging to calculate, because different seeding impurities have been adopted in different simulations. However, figure \ref{fig:global_imps} (b) shows that for the same seeded species the peaking does not show a strong trend with $Z_{eff}$, while lower values are found employing Ne instead of Ar, suggesting that other effects, like the species mass, may play a crucial role in the transport predictions. \\
The radial concentrations of H, W and seeding species found across the scans described in the previous section have been analyzed to quantify core impurity peaking with respect to the electrons. The profiles are shown in figure \ref{fig:imp_conc}. Mostly flat concentrations are found for W / H (a) and seeding species (b), validating the approach of prescribing fixed radial concentrations used in [Howard, JPP 2026, accepted]. A similar result was found for SPARC in \cite{muraca_impurity_2026}. $f_H$ reaches $\sim5-6\%$ concentrations on axis, suggesting that ICRH minority heating should be efficiently absorbed in the core. In the right plot, only the profiles from simulations where $\epsilon$ has been calculated with Eq. \ref{eq:Kallenbach_reg} are shown, excluding enrichment scans, in order to highlight better the shape of the profiles. The results from the $\epsilon$ scans show similar features. \\
\begin{figure}[h]
    \centering
    \includegraphics[width=0.49\linewidth]{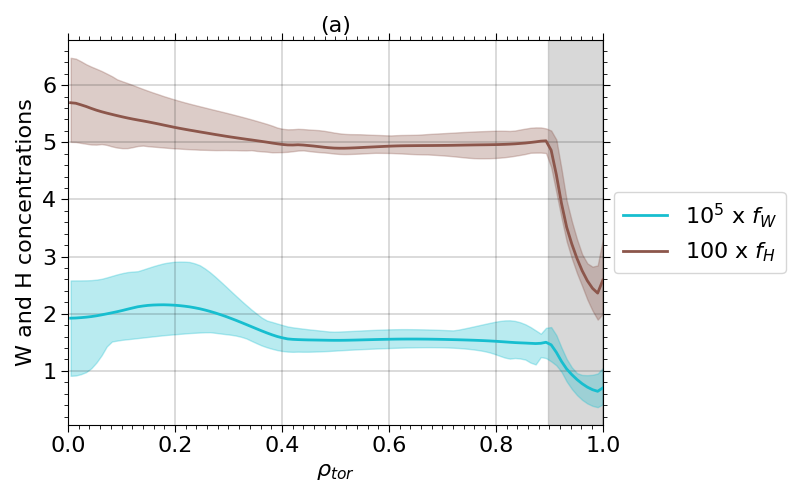}
    \includegraphics[width=0.49\linewidth]{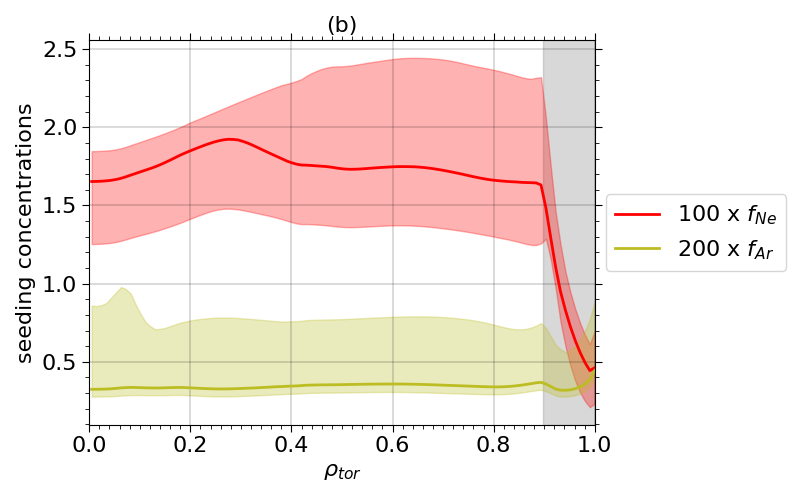}
    \caption{Left: W (blue) and H (brown) radial concentration profiles for all the scans (i.e. points in figure \ref{fig:global_imps}); right: Ne (red) and Ar (green) radial concentration profiles for scans in $n_{sep}/n_{ped}$ and $n_{ped}$ at fixed $\epsilon$ (computed with Eq. \ref{eq:Kallenbach_reg}). The lines indicate average profiles, while the shaded areas show the minimum and maximum values across scans. The gray vertical shade highlights the pedestal region. $\rho_{tor}$ is the square root of the normalized toroidal flux.}
    \label{fig:imp_conc}
\end{figure}
In summary, this section shows that in the core the turbulent impurity transport prevails on the neoclassical component, predicting moderate high-Z impurity peaking and confirming that roughly constant concentrations are likely a reasonable approximation. The low impurity peaking is consistent with predictions for ITER \cite{fajardo_theorybased_2025}.

\subsection{Effect of rotation}
The simulations in the previous sections show turbulent dominance for impurity transport in the core. However, null toroidal velocity has been considered. This assumption usually causes higher turbulent transport, but can exhibit low impurity accumulation in the core \cite{angioni_neoclassical_2014, fajardo_analytical_2023}. For this reason, momentum transport and rotation predictions have been conducted for the nominal Ar-seeded case, coupling the analytical model developed in \cite{zimmermann_experimental_2024} with ASTRA, and including its effect on impurity transport through the Mach number and radial electric field.
\begin{figure}[h]
    \centering
    \includegraphics[width=0.48\linewidth]{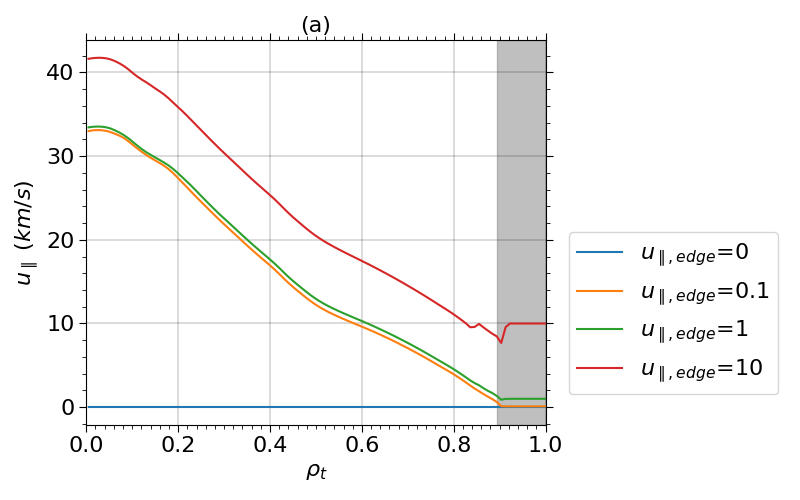}
    \includegraphics[width=0.48\linewidth]{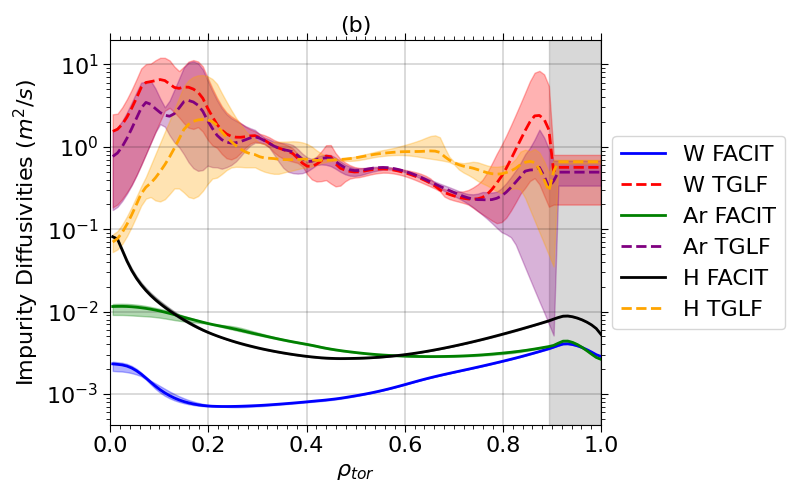}
    \caption{Profiles of toroidal velocity (a) and impurity diffusivity for all the species (b) for a scan in edge toroidal velocity. On the right plot, the lines indicate average profiles, while the colored areas show the maximum and minimum values. The gray zone represents the pedestal region.}
    \label{fig:rotation}
\end{figure}
This model, which provides momentum diffusivity, convection and intrinsic torque through analytical expressions, was validated on AUG and DIII-D discharges \cite{zimmermann_analysis_2022, zimmermann_experimental_2024}, featuring toroidal velocities between 25 and 125 $km/s$. The model does not include the effect of NTV torque \cite{clement_neoclassical_2022, park_selfconsistent_2017} and ripples \cite{fenzi_plasma_2011}, whose description is beyond the scope of the present article. Since the model was validated in the core, a flat toroidal rotation has been assumed from top of pedestal to the separatrix and scanned from 0.1 to 10 $km/s$ to take uncertainties into account. This choice is supported by \cite{mcdermott_edge_2009}, which showed a weak variation of the toroidal rotation in the edge of ICRH-dominated plasmas for Alcator C-mod. Since the momentum transport in ASTRA is solved for the velocity parallel to the magnetic field lines (i.e. $u_\parallel$), the assumption $v_{tor}=u_{\parallel}$ has been done, where $v_{tor}$ is the toroidal velocity. This assumption holds for low poloidal rotation, which aligns with recent findings for AUG \cite{lebschy_measurement_2018}. The variation of kinetic profiles and fusion power is $<1\%$ across the scan.
\begin{figure}[h]
    \centering
    \includegraphics[width=0.48\linewidth]{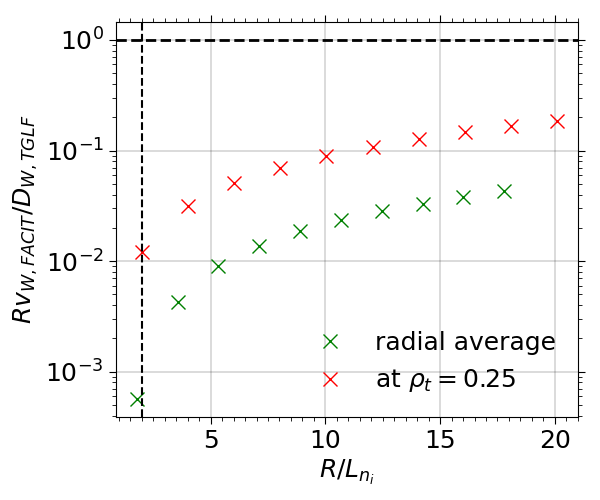}
    \includegraphics[width=0.48\linewidth]{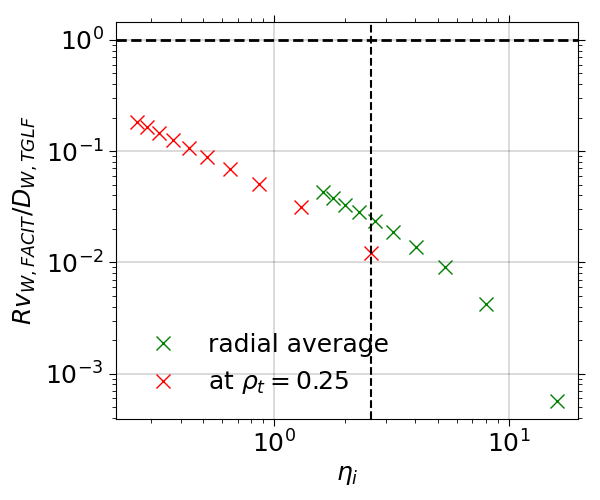}
    \includegraphics[width=0.48\linewidth]{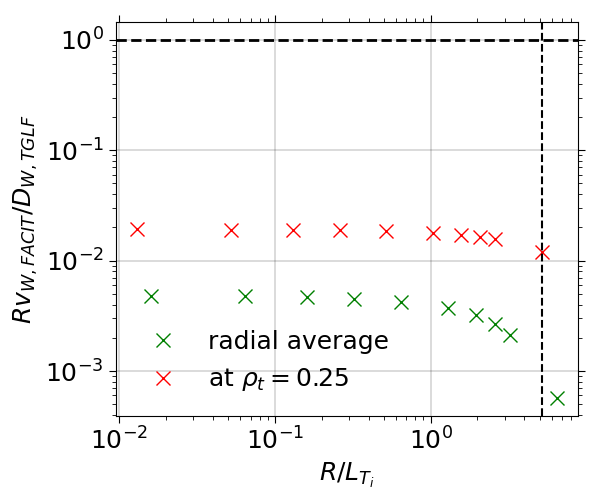}
    \includegraphics[width=0.48\linewidth]{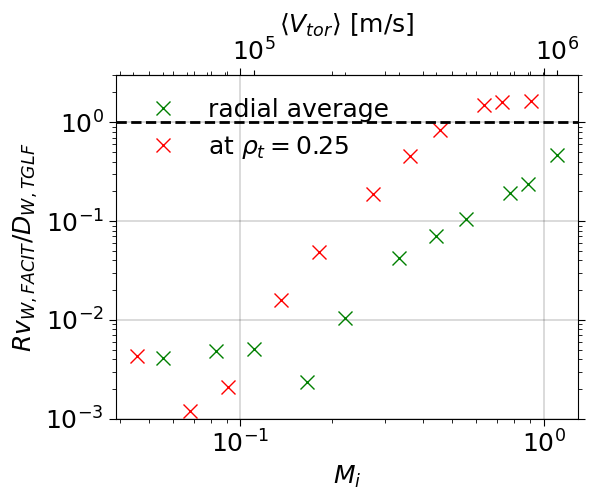}
    \caption{Ratios between ($R\cdot$) W neoclassical convection and turbulent diffusivity for several scans: the top plots show a scan in ion density gradient; the bottom left plot indicates a scan in ion temperature gradient; the bottom right plot highlights a scan in rotation. The green crosses indicate average values, while the red ones are at $\rho_t=0.25$. $\eta_i$ is the ratio between normalized ion temperature and density gradients. The vertical dashed lines indicate the nominal conditions (not shown for the toroidal velocity scan, where the nominal rotation is null).}
    \label{fig:FACIT_scans}
\end{figure}
The variation of $\langle \left| v_{NC} \right| /(\left| v_{NC} \right|+\left| v_{turb} \right|)\rangle$ is maximum 8\% for both W and Ar. Instead, $\langle D_{NC}/(D_{NC}+D_{turb})\rangle$ shows higher variability, reaching 17 and 160 \% higher values increasing rotation for W and Ar respectively. However, since the neoclassical to total diffusivity ratio remains quite low for all cases ($\langle D_{NC}/D_{tot}\rangle<2\%$), the effect on the impurity profiles is negligible. In particular, the W and Ar density peaking variation is respectively 3 and 2 \%.
The profiles of toroidal velocity and W neoclassical / turbulent diffusivity are shown in figure \ref{fig:rotation}. The W convection profiles are similar to those shown in figure \ref{fig:profiles_nscan} (d).
One can notice that even when assuming edge velocity 10 $km/s$, the core reaches maximum 40 $km/s$. The radially averaged toroidal velocity is roughly $\sim20\:km/s$, in agreement with simple estimates from \cite{parra_scaling_2012}, \cite{rice_impurity_2000} and \cite{rice_observations_2001}.  \\
The low neoclassical to total diffusivity ratio is justified by the low plasma collisionality. In fact, in such conditions, changing rotation only weakly affects the neoclassical transport \cite{fajardo_analytical_2023}. In order to further strengthen the robustness of this finding, a deeper analysis has been conducted using FACIT standalone. In particular, ion density gradient, ion temperature gradient and rotation scans have been performed, since these are considered the main parameters affecting the impurity neoclassical transport \cite{field_peripheral_2023, fulop_nonlinear_1999, fajardo_analytical_2023}. The ratios between the neoclassical convection resulting from the scans, multiplied by the major radius, and the turbulent diffusivity from the nominal ASTRA+TGLF simulation are shown in figure \ref{fig:FACIT_scans}, inspired by figure 11 of \cite{fajardo_theorybased_2025}. The scan in density gradient (upper plots) shows that even with unphysical $R/L_{n_i}$, $R\cdot V_{FACIT}/D_{TGLF}$ reaches 0.2. The $R/L_{T_i}$ scan shows less than 2\% neoclassical contribution, even if nearly-flat temperature profiles are assumed. Finally, the rotation scan exhibits a high neoclassical transport percentage (i.e. $>50\%$) only for Mach numbers higher than 0.4 (or toroidal velocities above $300 \: km/s$), which are not consistent with the results found in the simulations with momentum transport. \\
Therefore, turbulence robustly dominates transport for impurities in the core and the presence of rotation does not sensitively affect the predicted W density peaking.

\section{Conclusions}
In this article, integrated modeling of ARC H-modes has been conducted, examining the feasibility of high performing detached scenarios. The results have shown that fusion power values around 950 MW can be reached, ensuring 2 eV temperatures at the divertor, by adopting Ar as seeded impurity. While an $n_{ped}$ scan has shown small impact on the results, sensitivity scans of impurity enrichment and separatrix density have shown a non-negligible effect on fusion power, with values varying between 750 and 1000 MW. $n_{sep}/n_{ped}$ has been found to impact the performance in a non-negligible way through its modification of the peeling-ballooning stability curve, leading to a 40 kPa drop in the pedestal pressure, when increasing $n_{sep}/n_{ped}$ from 0.4 to 0.5. $T_{sep}$ has been found to vary roughly between 300 and 400 eV. \\
Scenarios with Ne seeding have also been tested, scanning the same input parameters. In this work, Ar is found to be the best solution to achieve high performance while detaching the divertor, always exhibiting robust H-mode access according to the Schmidtmayr scaling, while Ne-seeded plasmas have shown lower fusion power ($<800\:MW$) and less robust H-mode conditions. The reason behind the lower performance is the low enrichment factor predicted with Eq. \ref{eq:Kallenbach_reg}, and the consequent excessive Ne core accumulation, which causes lower DT concentration and high radiation. A decreasing trend of performance with $n_{sep}/n_{ped}$ is confirmed also by Ne simulations. \\
The turbulence prevails on the neoclassical component for W and Ar/Ne transport in the core. This results in roughly flat concentration profiles and low W density peaking, with decreasing values at higher $Z_{eff}$. The H concentration near-axis is $\sim 5-6\%$, showing values compatible with ICRH absorption. \\
A reduced model for momentum transport \cite{zimmermann_experimental_2024} has been coupled to the framework to provide a rotation estimate for several edge boundary conditions, slightly changing impurity profiles, but not affecting the fusion performance, exhaust conditions and turbulent / neoclassical transport components; a sensitivity study of neoclassical W transport changing $\nabla n_i$, $\nabla T_i$ and rotation show that turbulence constantly dominates particle impurity transport. \\
The results presented in this article support the feasibility of accessing H-mode and high performance in ARC, with realistic modeling of impurities and ensuring the protection of the divertor, through seeding of Ar/Ne and detachment.

\newpage
\section*{Acknowledgments}
The authors thank the MIT PSFC for its constructive feedback, in particular the MFE Integrated Modeling group.
A special thank to Enrico Panontin, Gabriele Ferrero and Paola Muscente for interesting discussions and support.
The authors acknowledge the use of ChatGPT during the article editing phase. This research used resources of the National Energy Research Scientific Computing Center, a DOE Office of Science User Facility using NERSC award FES-ERCAP0032161, for the EPED simulations used to train the neural nework model. The ASTRA simulations (ASTRA from main branch with hash a00f496a5489e12bbdbdc02dc38482057dd43b0b) presented in this paper were performed on the MIT-PSFC partition of the Engaging cluster at the MGHPCC facility (www.mghpcc.org) which was funded by DoE grant number DE-FG02-91-ER54109. Davide Silvagni is funded within the framework of the EUROfusion Consortium, funded by the European Union via the Euratom Research and Training Programme (Grant Agreement No 101052200 — EUROfusion). Views and opinions expressed are however those of the author(s) only and do not necessarily reflect those of the European Union or the European Commission. Neither the European Union nor the European Commission can be held responsible for them. \\
\textit{This work was supported by CFS under RPP020 fundings}.\\

\printbibliography

\end{document}